\documentclass[11pt]{article}
\usepackage[utf8]{inputenc}
\usepackage{geometry}
\usepackage{amsmath, amsthm, amssymb, graphicx}
\usepackage{graphicx}
\usepackage[parfill]{parskip}
\usepackage{booktabs}
\usepackage{array}
\usepackage{paralist}
\usepackage{verbatim}
\usepackage{subfig}
\usepackage[nottoc,notlof,notlot]{tocbibind}
\usepackage[titles,subfigure]{tocloft}
\theoremstyle{definition}

\theoremstyle{plain}

\theoremstyle{remark}
\newtheorem{rem}{Remark}[section]

\newcommand{\bena}{\begin{eqnarray}\begin{array}{l}}
		\newcommand{\eena}{\end{array}\end{eqnarray}}
\newcommand{\ben}{\begin{eqnarray}}
	\newcommand{\een}{\end{eqnarray}}
\newcommand{\bea}{\begin{array}}
	\newcommand{\eea}{\end{array}}
\newcommand{\bes}{\begin{subequations}}
	\newcommand{\ees}{\end{subequations}}
\newcommand{\bec}{\begin{cases}}
	\newcommand{\eec}{\end{cases}}
\newcommand{\bef}{\begin{figure}[H]}
	\newcommand{\eef}{\end{figure}}
\newcommand{\bet}{\begin{tikzpicture}}
	\newcommand{\eet}{\end{tikzpicture}}
\newcommand{\beq}{\begin{equation}}
	\newcommand{\eeq}{\end{equation}}
\newcommand{\bep}{\begin{proof}}
	\newcommand{\eep}{\end{proof}}
\def\beg#1\eeg{\begin{align}#1\end{align}}
\def\besl#1\eesl{\begin{subequations}\begin{align}#1\end{align}\end{subequations}}

\newcommand{\parl}[2]{\ensuremath{\frac{\partial #1}{\partial #2}}}

\newcommand{\abs}[1]{\lvert{#1}\rvert}
\def\inc(#1){\includegraphics[height=3 cm]{pics/#1}}

\def\bC{\ensuremath{{\bf C}}}
\def\bF{\ensuremath{{\bf F}}}

\def\bp{\ensuremath{{\bf p}}}
\def\bx{\ensuremath{{\bf x}}}

\def\bv{\ensuremath{{\bf v}}}
\def\bM{\ensuremath{{\bf M}}}
\def\bL{\ensuremath{{\bf L}}}

\def\bH{\ensuremath{{\bf H}}}
\def\bX{\ensuremath{{\bf X}}}
\def\bI{\ensuremath{{\bf I}}}
\def\bD{\ensuremath{{\bf D}}}

\def\bR{\ensuremath{{\bf R}}}
\def\br{\ensuremath{{\bf r}}}
\def\bp{\ensuremath{{\bf p}}}
\def\bC{\ensuremath{{\bf C}}}

\begin{document}
\title{A Thermodynamically Consistent Model for Yield Stress Fluids}
\author{ Nan Jiang \footnote{Beijing Computational Science Research Center, Beijing, 100193, China} and Qi Wang \footnote{Department of Mathematics, University of South Carolina, Columbia, SC 29208, USA}}

\date{}
	
\maketitle
\begin{abstract}
In this study, we formulate a thermodynamically consistent rheological model for yield stress fluids by introducing an internal dynamic variable and extending the framework established by Kamani et al (2021) and the classical Oldroyd-B model. The dynamics of the internal variable capture the material's transient response to changes in deformation, characterized by an effective relaxation time, elastic modulus, and viscosity. To assess the model's validity and range of applicability, we compare it with the recently developed Kamani-Donley-Rogers (KDR) model in terms of various material and rheometric functions, highlighting both divergences and parallels between the two models. Our numerical results on a host of material functions and rheological parameters illustrate the practical applicability and advantages of the new thermodynamically consistent model over the KDR model. Specifically, the new model complies with the second law of thermodynamics and can describe a broader range of rheological properties of yield stress fluids.


\end{abstract}
\section{Introduction}

\noindent \indent The yield stress fluid represents a distinct subset of non-Newtonian fluids characterized by their unique material behavior. This fluid remains in a solid-like state, impervious to macroscopic flow, when subjected to shear stress below a specific threshold value. However, it transitions into a fluid-like state only when the applied shear stress surpasses this critical threshold, referred to as the material's yield stress (\cite{Barnes2007,RMP,Coussot2006,Coussot2019,Moller2006}). Yield stress fluids with the same composition can thus exhibit distinct responses under different stress conditions.
Encountered in various daily products such as toothpaste, pigments, and creams, yield stress fluids play a ubiquitous role in everyday life. Their application extends across diverse industrial sectors, including food, chemicals, and construction, owing to the distinctive rheological properties they exhibit.

The viscosity of yield stress fluids can be broadly categorized based on its dependence on shear rate. This leads to two main classifications: simple yield stress fluids(\cite{Bertola2003,Moller2009,Ovarlez2013,Becu2006}) and thixotropic yield stress fluids(\cite{Moller2009,Moller2008,Coussot2007}). In the case of simple yield stress fluids, such as Carbopol gel, viscosity remains unaffected by changes in strain rate over time. Conversely, thixotropic yield stress fluids, exemplified by bentonite, display viscosity variations contingent on the duration of applied shear force. The distinction between these categories can be ascertained through up-down stress sweeps experiments, wherein stress hysteresis becomes observable for thixotropic yield stress fluids. These experiments prove valuable in determining whether a material exhibits thixotropic behavior.
The versatile nature of yield stress fluids, transitioning between solid and fluid states based on applied shear stress, makes them useful to numerous applications. It is therefore important to understand and categorize their rheological properties.

The earlier constitutive models to describe the yield stress fluid is the Bingham model \cite{Bingham}. The Bingham model is the simplest constitutive  equation for describing the stress-strain relation of a fluid under yield:
\ben
\left \{
\bea{ll}
\sigma=\sigma_{y}+\eta_{p}\dot{\gamma}, &\sigma>\sigma_{y}\\
\gamma=0, & \sigma<\sigma_y,
\eea\right.
\een
where $\sigma$ is the shear stress, $\sigma_{y}$ is the yield stress, $\eta_{p}$ is a constant viscosity, and $\dot{\gamma}$ is the shear rate. It shows that the material exhibits rheological properties of Newtonian fluids after yielding whereas its deformation is zero before yielding. The other two commonly used models for yield stress fluids are the Herschel-Bulkley model and the Casson model\cite{Casson}, respectively. They differ from the Bingham model when the stress is beyond the yield stress:
\ben
\bea{l}Herschel-Bulkley: \sigma=\sigma_{y}+k\dot{\gamma}^n, \sigma>\sigma_{y};\\\\
 Casson: \sigma^{\frac{1}{2}}=\sigma_{y}^{\frac{1}{2}}+(\eta_{p}\dot{\gamma})^{\frac{1}{2}}, \sigma>\sigma_{y};.
 \eea\een
Here $k$,$n$ are constant parameters. The Bingham model is a special case of the Herschel-Bulkley model with $n=1$. The Herschel-Bulkley model fits  the rheological behavior of polymer materials well such as gels and suspensions. The Casson model gives a good description of the rheological behavior of blood. These constitutive equations only describe the fluid behavior without addressing the transition from the solid to fluid when the stress varies across the yield stress continuously.

Note that yield stress fluids maintain their composition despite variations in applied stress and it is the material's mechanical response that exhibits distinct behavior at different stress regimes. Therefore, it is important to emphasize the importance of developing constitutive equations that not only capture rheological transitions but also elucidate the mechanisms underlying the generation of yield in modeling the yield stress fluids. The constitutive relationships between shear stress and shear rate provided by existing models are primarily designed for steady-state shear scenarios. Unfortunately, these models fall short in explaining the dynamic transition of yield stress fluids from a solid to a liquid state.

There exist three primary mechanisms to explain the yielding of materials: glass transition, jamming transition, and gelation. The glass transition is linked to restricting the Brownian motion of small colloidal particles(\cite{Royall2013,Ikeda2012,Ileda2013,Megen1989}), while the jamming transition obstructs the motion between larger colloidal particles(\cite{Ikeda2012,Ileda2013,Liu2010}). Both transitions are contingent on the volume fraction $\phi$ occupied by colloidal particles. Critical values $\phi_{g}$ and $\phi_{j}$ correspond to different viewpoints. Specifically, when $\phi>\phi_{g}$ (or $\phi>\phi_{j}$), flow curves conform well to the Herschel-Bulkley model with yield stress. Conversely, when $\phi<\phi_{g}$ (or $\phi<\phi_{j}$), the rheological properties resemble those of Newtonian fluids without yield stress. Gelation involves the formation of a stable network between colloidal particles through attractive forces, resulting in the manifestation of yield stress (\cite{Royall2013,Larson1999,Zaccarelli_2007}). These mechanisms elucidate how meso-structures can contribute to the distinctive rheological properties exhibited by yield stress fluids.

Recently, Krutarth Kamani, Gavin J. Donley, and Simon A. Rogers proposed a macroscopic/continuum constitutive equation  for simple yield stress fluids based on the classical Oldroyd-B model \cite{Rogers2021}, given by
\beq
\sigma+\frac{\eta_{p}+\eta_{s}}{G}\check{\sigma}=2\eta_{p}(\bD+\frac{\eta_{s}}{G}\check{\bD}),\label{Rogers}
\eeq
where $\bf\sigma$ is the extra stress tensor, $\bD$ is the strain rate tensor, $\bf\check{\sigma}$ and $\bf\check{D}$ are the upper-convected time derivatives of $\bf\sigma$ and $\bD$, respectively, $\eta_{s}$ and $G$ are the viscosity and elastic modulus in the fluid regime, $\eta_{p}$ is the Herschel-Bulkley viscosity defined by
\ben
\bea{l}
\eta_{p}=\sigma_{y}/\|\dot{\gamma}\|+k\|\dot{\gamma}\|^{n-1}=\frac{1}{\dot{\gamma}}(\sigma_y+k\|\dot{\gamma}\|^n),
\eea
\een
$\dot{\gamma}$ is the second invariant of the rate of strain tensor,
$
\|\dot{\gamma}\|=\sqrt{2\bD:\bD-2tr(\bD)^{2}},
$
$\sigma_{y}$ is the yield stress, $k$ and $n$ are constant parameters in the model. We refer to this model as the KDR model in this paper \cite{Rogers2021}.
This constitutive equation reduces to the Oldroyd-B model when $\dot{\gamma}$ is assumed a constant. It is known that the Oldroyd-B model is thermodynamically consistent, however, the KDR model is not.

The objective of this paper is to develop a thermodynamically consistent model for yield stress fluids, extending both the Oldroyd-B model and the KDR model. Our strategy is to introduce an internal dynamic variable  $\lambda$ to describe the mesoscopic structural evolution in yield stress fluids. Then, we postulate its dynamics in such a way that its asymptotic limit yields the materials parameters in the constitutive equation of the KDR model (\ref{Rogers}). We then show rigorously that this new model is thermodynamically consistent and compare it with the KDR model with respect to a host materials functions and rheometric functions. After an extensive comparison, we conclude that the new thermodynamically consistent model (TC model) is more versatile in terms capturing various rheological behavior of the yield stress fluids in all simple rheological flows.

The rest of the paper is organized as follows. In \S 2, we briefly review the thermodynamical consistency of the classical Oldroyd-B model and derive the new thermodynamically consistent model for yield stress fluids. In \S3, we assess the thermodynamically consistent constitutive model through various materials functions and rheometric functions  in simple flows in contrast to the KDR model. We conclude the study in \S4.

\section{Thermodynamical consistent yield stress model }

\noindent \indent We recall briefly the classical Oldroyd-B model and its derivation from the perspective of the generalized Onsager principle \cite{Bird1987,Larson1999,Wang2021,Masao2018}. Then, we formulate a new thermodynamically consistent yield stress model based on an extended framework.

\subsection{Thermodynamical consistent formulation of the Oldroyd-B model}
\noindent \indent  We assume that the free energy density of the complex fluid system is a functional of  the positive definite conformation tensor $\bC>0$. The Helmholtz free energy of the fluid system is given by\cite{Masao2018}
\beq
A=\int\frac{1}{2}Gtr(\bC-\ln\bC-\bI)d\bx,
\eeq
where $G$ is the elastic modulus of the complex fluid.
The time rate of change of the Helmholtz free energy is given by
\beq
\frac{dA}{dt}=\int\frac{1}{2}G(\bI-\bC^{-1}):\frac{\partial \bC}{\partial t}d\bx=\int[\frac{1}{2}G(\bI-\bC^{-1}):\check{\bC}+G(\bC-\bI):\bD]d\bx,
\eeq
where
\ben
\check{\bC}=(\frac{\partial }{\partial t}+\bv\cdot \nabla) \bC-\bL\cdot \bC-\bC\cdot \bL^T
\een
is the upper-convected derivative (an invariant time derivative), $\bv$ is the mass average velocity,  and $\bL=\nabla \bv$ is the velocity gradient tensor.

We assume that the yield stress fluid is incompressible, the mass density is a constant and the temperature of the fliud is as constant as its surroundings. Therefore, the system does not absorb or release heat from the surroundings, nor does it do work on the outside. The momentum balance and incompressible equation in the Eulerian coordinate are given, respectively,  as follows:
\ben
\left \{
\bea{l}
\rho[\frac{\partial \mathbf{v}}{\partial t}+\mathbf{(v\cdot\nabla)v}]=\nabla (-p\bI+\sigma) +\bF_e,\\\\
\nabla \cdot \bv=0,
\eea\right.
\een
where $\rho$ is the fluid density,  $p$ is the hydrostatic pressure, $\sigma$ the extra stress tensor, and $\bF_e$ is the external force per unit volume. 

Accounting for momentum and mass conservation, the dissipation rate of the kinetic energy is calculated as follows
\beq
\begin{split}
\frac{dE_{k}}{dt}
&=\frac{d}{dt}\int\frac{1}{2}\rho \mathbf{v}^{2}dV=\int\frac{\partial}{\partial t}\frac{1}{2}\rho \mathbf{v}^{2}d\bx\\
&=-\int(\sigma:\bD+\bv\cdot\bF_{e})d\bx-\int\nabla\cdot[\rho\bv(\frac{\bv^{2}}{2}+\frac{p}{\rho})-\bv\cdot\sigma]d\bx.
\end{split}
\eeq
We define the total energy of the fluid system by
\ben
E=E_{k}+A=\int [\frac{1}{2}\rho \|\mathbf{v}\|^{2}+\frac{1}{2}Gtr(\bC-\ln\bC-\bI)]d\bx.
\een
We ignore the external force and the boundary term of the integral. Then the  dissipation rate of the energy is given by
\beq
\frac{dE}{dt}=-\int\{[\sigma-G(\bC-\bI)]:\bD-\frac{1}{2}G(\bI-\bC^{-1}):\check{\bC}\} d\bx.
\eeq
Recall the first law of thermodynamics:
\beq
dE_{k}+dU=dQ-dW
\eeq
Where $dU$ is the increase of the internal energy of the matter system, $dQ$ is the heat added to the system and $dW$ is the work done to the surrounding by the system. Here we have:
\beq
dQ=dW=0,\ dU=dA+TdS+SdT=dA+TdS
\eeq
Where $T$ is the temperature and $S$ is the entropy of the system. So we get
\beq
\frac{dE}{dt}=-T\frac{dS}{dt}
\eeq
The Onsager principle\cite{Wang2021} gives 
\beq
\frac{dS}{dt}=\dot{\bx}^{T}\bX=\dot{\bx}^{T}\bR\dot{\bx}
\eeq
Where $\bR$ is the friction operator/matrix which is symmetric and $\dot{\bx}$ is the time derivative of the thermodynamic variable $\bx$ which used to describe the system, $\bX=\parl{S}{x}$ is the thermodynamic force. It implies that there exists the dynamics equation:
\beq
\dot{\bx}=\bR^{-1}\bX
\eeq
In fact, GENERIC\cite{Esen2022,Grmela2018} can give a more universal formalism. Part of the energy of the system is conserved, and we denote this part by the Hamiltonian $\bH$. Then the dynamic equation should be
\beq
\dot{\bx}=\{\bx,\bH\}_{f(\br,\bp)}+\bR^{-1}\bX\label{IRE}
\eeq
\beq
\{\bx,\bH\}_{f(\br,\bp)}=\int d\br\int d\bp f(\br,\bp)(\parl{\bx}{\br}\parl{\bH}{\bp}-\parl{\bH}{\br}\parl{\bx}{\bp})
\eeq
Where $(\br,\bp)$ is the coordinates in phase space and $f(\br,\bp)$ is the distribution function in phase space. But the first term on the right-hand side of equation(\ref{IRE}) is difficult to obtain. \par
Here we conjecture a constitutive equation for the complex fluid system based on the Onsager principle as follows
\beq
\begin{pmatrix}
	\sigma-M(\bC-I)\\
	-\frac{1}{2}M(I-\bC^{-1})
\end{pmatrix}=\begin{pmatrix}
2\eta\bI&0\\
0&\frac{1}{2}M\tau\bC^{-1}
\end{pmatrix}\cdot \begin{pmatrix}
\bD\\
\check{\bC}\end{pmatrix},
\eeq
where $\tau$ is the relaxation time and $\eta$ is the fluid viscosity. This is a simple constitutive equation with the friction operator/matrix in a diagonal form.
It implies
\beq
\left \{
\bea{l}
\sigma-M(\bC-\bI)=2\eta\bD,
\\\\
\check{\bC}=-\frac{1}{\tau}(\bC-\bI).
\eea\right.
\eeq
Combining the above, we arrive at the classical Oldroyd-B constitutive equation
\beq
\sigma+\tau\check{\sigma}=2(\eta+M\tau)\bD+2\eta\tau\check\bD,
\eeq
where $M,\tau,\eta$ are constant parameters. 

We can decompose the extra stress tensor into the viscous part and viscoelastic part as follows
\ben
\sigma=2\eta \bD+\sigma_e, \quad \sigma_e=M(\bC-\bI),
\een
where $\sigma_e$ is the viscoelastic stress tensor. The Oldroyd-B model can be written into
\beq
\sigma_e+\tau\check{\sigma}_e=2\eta_e\bD,
\eeq
where $\eta_e=M\tau$ is a viscosity parameter.

\begin{rem}
A general constitutive equation can be proposed as follows
\beq
\begin{pmatrix}
	\sigma-M(\bC-I)\\
	-\frac{1}{2}M(I-\bC^{-1})
\end{pmatrix}=\begin{pmatrix}
R_{11}&R_{12}\\
R_{12}&R_{22}
\end{pmatrix}\cdot \begin{pmatrix}
\bD\\
\check{\bC}\end{pmatrix},
\eeq
where $\bR=(R_{ij})$ is the friction operator/matrix and $\bM=\bR^{-1}$ is the mobility operator. Requirement $\bM\geq 0$ is the consequence of the  Onsager principle  or equivalently, the second law of thermodynamics \cite{Wang2020}.
\end{rem}

\subsection{Thermodynamically consistent yield stress model}

\noindent \indent The Kamani-Donley-Rogers (KDR) model for yield stress fluids [1] employs constitutive equation parameters that are prescribed as nonlinear functions of the second invariant of the strain rate tensor. To develop a thermodynamically consistent model, we introduce an internal dynamic variable, $\lambda$, and leverage the structure of the Oldroyd-B model, making the model parameters prescribed functions of this internal variable. Many papers have used similar methods such as\cite For the internal variable, we derive a dynamical model such that the model parameters asymptotically approach those in the KDR model (Equation \ref{Rogers}) as $t \rightarrow \infty$. We then demonstrate that the new model derived in this manner is thermodynamically consistent.

We consider the general case, where the parameters $M, \eta, \tau$ are variables that depend on $\lambda$. We adopt the total energy of the fluid system as follows
\ben
E=\int [\frac{1}{2}\rho \|\mathbf{v}\|^{2}+\frac{1}{2}Mtr(\bC-\ln\bC-\bI)]d\bx,
\een

 The energy dissipation of the system is calculated as follows
\beq
\frac{dE}{dt}=-\int\{[\sigma-M(\bC-\bI)]:\bD-\frac{1}{2}M(\bI-\bC^{-1}):\check{\bC}-\frac{1}{2}\dot{M}tr(\bC-\ln\bC-\bI)\} d\bx
\eeq

We apply the Onsager principle in the generalized Onsager principle \cite{Wang2020}. A constitutive equation for stress $\sigma$ is obtained as follows
\beq
\left \{
\bea{l}
\sigma-M(\bC-\bI)=2\eta\bD,
\\\\
\check{\bC}=-\frac{1}{\tau}(\bC-\bI).
\eea\right.\label{const-2}
\eeq
It reduces to
\beq
(\frac{1}{\tau}-\frac{\dot{M}}{M})\sigma+\check{\sigma}=2[(\frac{1}{\tau}-\frac{\dot{M}}{M})\eta+M+\dot{\eta}]\bD+2\eta\check\bD,\label{xxx}
\eeq
With \eqref{const-2}, the  energy dissipation rate of the model is calculated as follows
\beq
\frac{dE}{dt}=-\int\{2\eta\bD:\bD+\frac{M}{2\tau}tr(\bC+\bC^{-1}-2\bI)-\frac{1}{2}\dot{M}tr(\bC-\ln\bC-\bI)\} d\bx.
\eeq

For the system  to be energy dissipative, the following inequality must hold
\beq
\frac{M}{2\tau}tr(\bC+\bC^{-1}-2\bI)-\frac{1}{2}\dot{M}tr(\bC-\ln\bC-\bI)>0.
\eeq
Given that $\bC$ is positive definite  with eigenvalues $a_{i}(i=1,2,3)$, we have
\beq
\left \{
\bea{l}
tr(\bC+\bC^{-1}-2\bI)=\sum_{i=1}^{3}(a_{i}+a_{i}^{-1}-2),\\\\
tr(\bC-\ln\bC-\bI)=\sum_{i=1}^{3}(a_{i}-\ln a_{i}-1).
\eea\right.
\eeq
Note that
\beq
a_{i}^{-1}=e^{-\ln a_{i}}\geq-\ln a_{i}+1.
\eeq
It follows that
\beq
tr(\bC+\bC^{-1}-2\bI)\geq tr(\bC-\ln\bC-\bI).
\eeq
A sufficient condition to ensure energy dissipation is given by
\beq
\frac{\dot{M}}{M}<\frac{1}{\tau}.
\eeq

We define the effective relaxation time 
\beq
\tau_{e}=\frac{\tau M}{M-\tau\dot{M}}>0
\eeq

Then the constitutive equation is
\beq
\sigma+\tau_{e}\check{\sigma}=2[\eta+\tau_{e}(M+\dot{\eta})]\bD+2\tau_{e}\eta\check\bD,\label{xxxx}
\eeq

We can get the KDR model from Equation (\ref{xxxx}) if the following asymptotic behavior exists
\ben
M\to(\frac{\eta_{p}}{\eta_{s}+\eta_{p}})^{2}G,\ \ 
\eta\to\frac{\eta_{p}\eta_{s}}{\eta_{s}+\eta_{p}},\ \ 
\tau_{e}\to\frac{\eta_{s}+\eta_{p}}{G}.
\een

There is a simplest dynamics that satisfies sunch an asymptotic relation
\ben
\bea{l}
\dot{\lambda}=\frac{G\alpha}{\eta_{s}}(g(\dot{\gamma})-\lambda),\quad
M=G\lambda^{2},\\
\eta=\eta_{s}\lambda,\quad
\tau_{e}=\frac{\eta_{s}}{G(1-\lambda)}.
\eea
\een

This dynamical system admits a unique stable steady $\lambda=g(\dot{\gamma})=\frac{\eta_{p}}{\eta_{s}+\eta_{p}}$  asymptotically as $t\to \infty$. Parameter  $\alpha >0$ can be used to adjust the relaxation time of the internal structural relaxation. 

When $\lambda$ reaches the steady state value, we have
\beg
\bea{l}
\lambda=\frac{\eta_{p}}{\eta_{s}+\eta_{p}}, \quad
M=\lambda^{2}G=(\frac{\eta_{p}}{\eta_{s}+\eta_{p}})^{2}G, \quad
\eta=\lambda\eta_{s}=\frac{\eta_{p}\eta_{s}}{\eta_{s}+\eta_{p}}, \quad
\tau_{e}=\frac{\eta_{s}}{G(1-\lambda)}=\frac{\eta_{p}+\eta_{s}}{G}.
\eea
\eeg
Plugging them into Equation (\ref{xxxx}), we obtain the KDR model (\ref{Rogers}). So, the KDR model is a model whose model parameters take on the asymptotic limits of the internal variable dynamics. Our model provides the missing temporal dynamics of the model parameters during the transient relaxation of the mesoscopic structure, rendering it thermodynamically consistent. 

In summary, the new constitutive model warrants a negative total energy dissipation rate. Thus, the resulting rheological model is thermodynamically consistent. In the viscoelastic stress $\sigma_e=M(\bC-\bI)$,
the constitutive equation reads as follows
\beq
(\frac{1}{\tau}-\frac{\dot{M}}{M})\sigma_e+\check{\sigma}_e=2M\bD.\label{xxx-2}
\eeq
The positivity of effective relaxation time $\tau_e=\frac{\tau M}{M-\tau \dot{M}}$ renders the thermodynamical consistency of the model. 
We next assess the thermodynamically consistent rheological model with respect to a plethora of material  and rheometric functions and contrast them with those of the KDR model's.

\section{Material functions in simple shearing and shear-free flows}
\noindent \indent
To assess the credibility and applicability of the yield stress fluid models it is imperative to evaluate their performance under controlled conditions, specifically in simple flows like shear and shear-free elongation. This evaluation encompasses an examination of material functions and crucial rheometric functions, including storage and loss modulus. In the subsequent sections, we present the results obtained from numerical experiments conducted in simple shearing flow and elongational flow, respectively.

\subsection{Material functions in simple shearing flows}

\noindent \indent We consider the yield stress fluid placed in a parallel plate rotational rheometer with a gap $h$. Let the angular velocity of the upper plate be $\Omega$ while the lower plate is fixed and the radius of the plates $R$. We measure the shear rate and shear stress of the fluid near the upper plate. The velocity gradient tensor in the cylindrical coordinates is given by
\ben
\bL=\dot{\gamma}\begin{pmatrix}
	0&0&0\\
	0&0&1\\
	0&0&0
\end{pmatrix}, \quad \dot{\gamma}=\frac{\partial\gamma}{\partial t}=\frac{\Omega r}{h}.
\een
The stress constitutive equation of (\ref{xxx}) in the component-wise form is given by
\beq
\begin{cases}
	(\frac{1}{\tau}-\frac{\dot{M}}{M})\sigma_{rr}+\dot{\sigma}_{rr}=0,\\
	(\frac{1}{\tau}-\frac{\dot{M}}{M})\sigma_{r\theta}+\dot{\sigma}_{r\theta}-\dot{\gamma}\sigma_{rz}=0,\\
	(\frac{1}{\tau}-\frac{\dot{M}}{M})\sigma_{rz}+\dot{\sigma}_{rz}=0,\\
	(\frac{1}{\tau}-\frac{\dot{M}}{M})\sigma_{\theta\theta}+\dot{\sigma}_{\theta\theta}-2\dot{\gamma}\sigma_{z\theta}=-2\eta\dot{\gamma}^{2},\\
	(\frac{1}{\tau}-\frac{\dot{M}}{M})\sigma_{z\theta}+\dot{\sigma}_{z\theta}-\dot{\gamma}\sigma_{zz}=[(\frac{1}{\tau}-\frac{\dot{M}}{M})\eta+M+\dot{\eta}]\dot{\gamma}+\eta\ddot{\gamma},\\
	(\frac{1}{\tau}-\frac{\dot{M}}{M})\sigma_{zz}+\dot{\sigma}_{zz}=0.
\end{cases}\label{xxx-1}
\eeq
We focus on shear stress $\sigma_{s}=\sigma_{z\theta}$ and the first normal stress difference, $\sigma_{N}=\sigma_{\theta\theta}-\sigma_{zz}$, here. If we set  $\sigma_{zz}=\sigma_{rr}=\sigma_{rz}=0$ initially, they remain zero for all time. Equation (\ref{xxx-1}) collapses to the following two equations
\beq
\left \{
\bea{l}
(\frac{1}{\tau}-\frac{\dot{M}}{M})\sigma_{N}+\dot{\sigma}_{N}-2\dot{\gamma}\sigma_{s}=-2\eta\dot{\gamma}^{2},\\\\
	(\frac{1}{\tau}-\frac{\dot{M}}{M})\sigma_{s}+\dot{\sigma}_{s}=[(\frac{1}{\tau}-\frac{\dot{M}}{M})\eta+M+\dot{\eta}]\dot{\gamma}+\eta\ddot{\gamma}.
\eea\right.
\eeq
Introducing a new quantity $\sigma_v$, we rewrite the above equations into the following
\beq
\bec
(1-\lambda)\sigma_{N}+\frac{\eta_{s}}{G}\dot{\sigma}_{N}=\frac{2\eta_{s}\dot{\gamma}}{G}(\sigma_{s}-\sigma_{v}),\\
(1-\lambda)\sigma_{s}+\frac{\eta_{s}}{G}\dot{\sigma}_{s}=\sigma_{v}+\frac{\eta_{s}}{G}\dot{\sigma}_{v},\\
\sigma_{v}=\lambda\eta_{s}\dot{\gamma},\\
\dot{\lambda}=\frac{\alpha G}{\eta_{s}}(g(\dot{\gamma})-\lambda).
\eec
\eeq
When the system approaches its asymptotic limit in a steady state as $t\to \infty$, we obtain
\beg
\lambda=g(\dot{\gamma})=\frac{\eta_{p}}{\eta_{s}+\eta_{p}}, \quad
\sigma_{v}=\lambda\eta_{s}\dot{\gamma}=\frac{\eta_{p}\eta_{s}}{\eta_{s}+\eta_{p}}\dot{\gamma}, \quad
\sigma_{s}=\eta_{p}\dot{\gamma}, \quad
\sigma_{N}=\frac{2\eta_{p}^{2}\dot{\gamma}^{2}}{G}=\frac{2\sigma_{s}^{2}}{G}.
\eeg

In the following, we compare responses of this new TC model with those of the KDR model in several simple shearing experiments.

\subsection{Step steady state flow}

\noindent \indent Experimentally, one often adjusts the shear rate of a sample to jump from one constant value to another to observe the stress response described by the model. Two representative experiments are the stress growth and the stress relaxation experiment, respectively, in which the shear rate of a sample varies from 0 to a constant value of $\dot{\gamma}$ and then from $\dot{\gamma}$ to 0.

Generally, we consider   a step-shear rate function of time as follows:
\beq
\dot{\gamma}=\dot{\gamma}^{-}+(\dot{\gamma}^{+}-\dot{\gamma}^{-})H(t)=
\bec
\dot{\gamma}^{-},\ t<0,\\
\dot{\gamma}^{+},\ t\geq0,
\eec
\eeq
where H(t) is the Heaviside or step function. We denote $f^{0}=f(0)$, $f^{+}=\lim\limits_{t\to0^{+}}f(t)$ and $f^{-}=\lim\limits_{t\to0^{-}}f(t)$ for any function $f(t)$.  It follows from equation(\ref{ge}) that
$\gamma,\sigma_{N},\lambda$ are continuous, whereas $\sigma_s$ is not with a jump discontinuity given by :
\begin{equation}
\sigma_{s}^{+}-\sigma_{s}^{-}=\sigma_{v}^{+}-\sigma_{v}^{-}=g(\dot{\gamma}^{-})\eta_{s}(\dot{\gamma}^{+}-\dot{\gamma}^{-})\label{mc}
\end{equation}
The shear stress undertakes a sudden change and the value before and after the change are given respectively by
\beq
\left \{
\bea{l}
\sigma_{s}^{-}=\eta_{p}^{-}\dot{\gamma}^{-},\\
\sigma_{s}^{0}=\sigma_{s}^{+}=\frac{\eta_{p}^{-}}{\eta_{p}^{-}+\eta_{s}}(\eta_{p}^{-}\dot{\gamma}^{-}+\eta_{s}\dot{\gamma}^{+}).
\eea\right.
\eeq
The dynamical equations of the system after the change in shear rate are given by
\beq
\bec
(1-\lambda)\sigma_{N}+\frac{\eta_{s}}{G}\dot{\sigma}_{N}=\frac{2\eta_{s}\dot{\gamma}}{G}(\sigma_{s}-\sigma_{v}),\\
(1-\lambda)\sigma_{s}+\frac{\eta_{s}}{G}\dot{\sigma}_{s}=\sigma_{v}+\frac{\eta_{s}}{G}\dot{\sigma}_{v},\\
\sigma_{v}=\lambda\eta_{s}\dot{\gamma}^{+},\\
\dot{\lambda}=\frac{\alpha G}{\eta_{s}}(g(\dot{\gamma}^{+})-\lambda).
\eec\label{ge}
\eeq
with initial conditions
\beg
\bea{l}
\lambda^{0}=1-g(\dot{\gamma}^{-})=\frac{\eta_{p}^{-}}{\eta_{s}+\eta_{p}^{-}}, \quad
\sigma_{v}^{0}=\lambda^{0}\eta_{s}\dot{\gamma}^{+}=\frac{\eta_{p}^{-}\eta_{s}}{\eta_{s}+\eta_{p}^{-}}\dot{\gamma}^{+}, \quad
\sigma_{N}^{0}=\frac{2}{G}(\eta_{p}^{-}\dot{\gamma}^{-})^{2}.
\eea
\eeg

In the KDR model,  the first normal stress difference $\sigma_{N}$ is continuous, while the jump of shear stress $\sigma_{s}$ is not given by
\beq
\sigma_{s}^{+}-\sigma_{s}^{-}=\int_{\dot{\gamma}^{-}}^{\dot{\gamma}^{+}}\frac{\eta_{s}\eta_{p}}{\eta_{p}+\eta_{s}}d\dot{\gamma}.
\eeq

\subsubsection{Stress growth upon inception of a steady shear flow}

\noindent \indent In a stress growth experiment, the sample changes abruptly from static to flowing at a constant shear rate $\dot{\gamma}$:
\ben
\dot{\gamma}(t)=
\left \{
\bea{ll}
0 & t<0,\\
\dot{\gamma} & t\geq 0.
\eea\right.
\een
 We calculate the relevant stress response from the following dynamical system.
\beq
\bec
(1-\lambda)\sigma_{N}+\frac{\eta_{s}}{G}\dot{\sigma}_{N}=\frac{2\eta_{s}\dot{\gamma}}{G}(\sigma_{s}-\sigma_{v}),\\
(1-\lambda)\sigma_{s}+\frac{\eta_{s}}{G}\dot{\sigma}_{s}=\sigma_{v}+\frac{\eta_{s}}{G}\dot{\sigma}_{v},\\
\sigma_{v}=\lambda\eta_{s}\dot{\gamma},\\
\dot{\lambda}=\frac{\alpha G}{\eta_{s}}(g(\dot{\gamma})-\lambda),
\eec\label{Incept-1}
\eeq
with initial  conditions
\beg
\lambda^{0}=1, \quad
\sigma_{v}^{0}=\eta_{s}\dot{\gamma}, \quad
\sigma_{s}^{0}=\eta_{s}\dot{\gamma}, \quad
\sigma_{N}^{0}=0.
\eeg

\begin{figure}[htbp]
		\centering
		\includegraphics[width=1\linewidth]{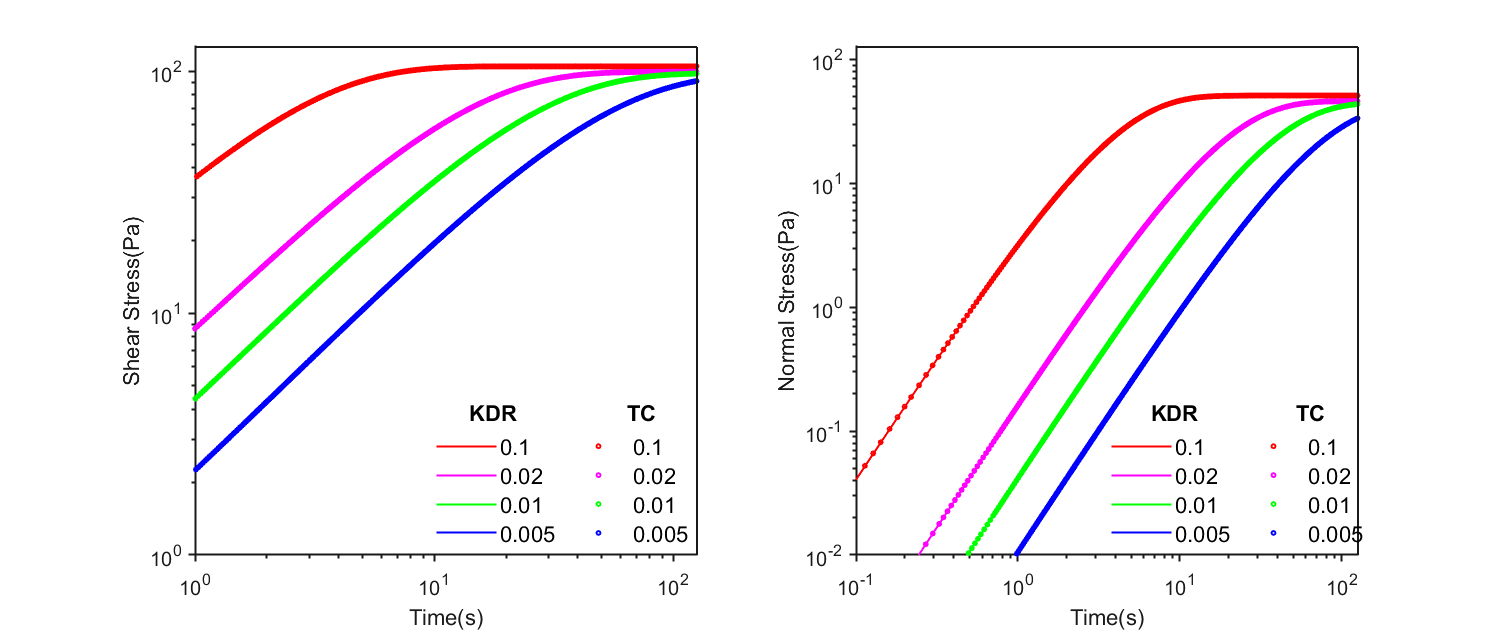}
		\caption[1]{Stress growth  for both the shear stress and the first normal stress of the KDR model and the new TC model in an inception of  steady shear flow  at shear rate 0.1, 0.02, 0.01, 0.005/s, respectively. The solid lines represent the results from the KDR  model and the dots represent those of the new TC model. The results obtained from the two different models are nearly indistinguishable in these experiments.
		}
\label{Stress-growth}
\end{figure}

For yield stress fluids, the shear stress and the first normal stress difference increase approximately linearly with respect to strain in the early stage of the steady shear experiment, similar to an elastic solid. After a long period of time, the stress reaches a plateau indicating the material yields and begins to flow, similar to a viscous fluid. This is shown in Figure \ref{Stress-growth} where  the KDR  model and the new  model are compared with each other and the results are the same quantitatively.

\subsubsection{Stress relaxation after cessation of a steady shear flow}

\noindent \indent In a stress relaxation experiment, the shear rate drops abruptly from a nonzero steady shear rate $\dot{\gamma}$ to zero
\ben
\dot{\gamma}(t)=\left\{
\bea{ll}
\dot{\gamma} & t<0,\\
0 & t\geq 0.
\eea\right.
\een
 The governing dynamical equations for the two pertinent stress functions are
\beq
\bec
(1-\lambda)\sigma_{N}+\frac{\eta_{s}}{G}\dot{\sigma}_{N}=0,\\
(1-\lambda)\sigma_{s}+\frac{\eta_{s}}{G}\dot{\sigma}_{s}=0,\\
\sigma_{v}=0,\\
\dot{\lambda}=\frac{\alpha G}{\eta_{s}}(1-\lambda),
\eec
\eeq
They imply
\beq
\frac{d}{dt}(\sigma_{s}e^{\frac{\lambda}{\alpha}})=\frac{d}{dt}(\sigma_{N}e^{\frac{\lambda}{\alpha}})=0, \quad \lambda>0.
\eeq
It indicates that the dynamics of the two rheometric functions are completed determined by the dynamics of the internal variable $\lambda$.
Using the initial conditions
\beg
\lambda^{0}=g(\dot{\gamma})=\frac{\eta_{p}}{\eta_{s}+\eta_{p}}, \quad
\sigma_{v}^{0}=0, \quad
\sigma_{s}^{0}=\frac{\eta_{p}^{2}\dot{\gamma}}{\eta_{s}+\eta_{p}}, \quad
\sigma_{N}^{0}=\frac{2}{G}(\eta_{p}\dot{\gamma})^{2},
\eeg
when the internal variable  reaches its steady state at $\lambda=1$,  we have
\beg
\sigma_{s}=\frac{\eta_{p}^{2}\dot{\gamma}}{\eta_{s}+\eta_{p}}e^{\frac{g(\dot{\gamma})-1}{\alpha}}, \quad
\sigma_{N}=\frac{2}{G}(\eta_{p}\dot{\gamma})^{2}e^{\frac{g(\dot{\gamma})-1}{\alpha}}.
\eeg
For the KDR model, the first normal stress doesn't even change after the shear rate suddenly  changes and the shear stress satisfies
\beq
\sigma_{s}=\eta_{p}\dot{\gamma}_{0}-\int^{\dot{\gamma}_{0}}_{0}\frac{\eta_{s}\eta_{p}}{\eta_{p}+\eta_{s}}d\dot{\gamma}.
\eeq

\begin{figure}[htbp]
		\centering
		\includegraphics[width=1\linewidth]{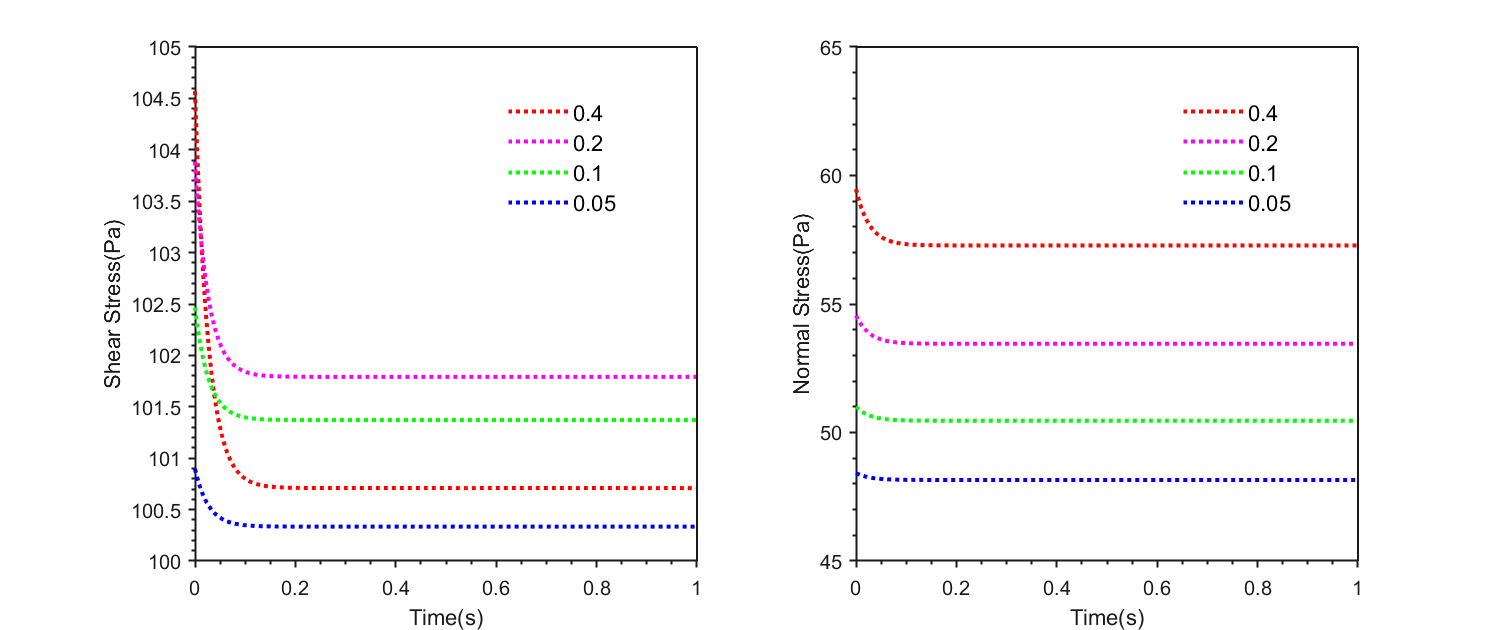}
		\caption[2]{Stress relaxation of the shear and the first normal stress difference of the new TC model. The stress decay to a priori determined values  with respect to shear rate 0.4, 0.2, 0.1, 0.05/s, respectively. }
\label{Stress-relax}
\end{figure}

It is perceived that the relaxed shear stress in a yield stress material would decay from its initial value to the yield stress of the material. From the formula of the shear stress in both models, we do see that at small $|\dot{\gamma}|<<1$ asymptotically. However, at a finite terminal shear rate $\dot{\gamma}$,  the TC model show the decay of the shear and the first normal stress difference to a specified stress value determined by the initial conditions instead of the yield stress. The final values may be higher or lower than the yield stress, dependent on the shear rate prior to the relaxation. The KDR model does not show any change of the shear stress in this experiment.
The results are depicted in Figure \ref{Stress-relax}.

\begin{figure}[htbp]
	\centering
	\includegraphics[width=1\linewidth]{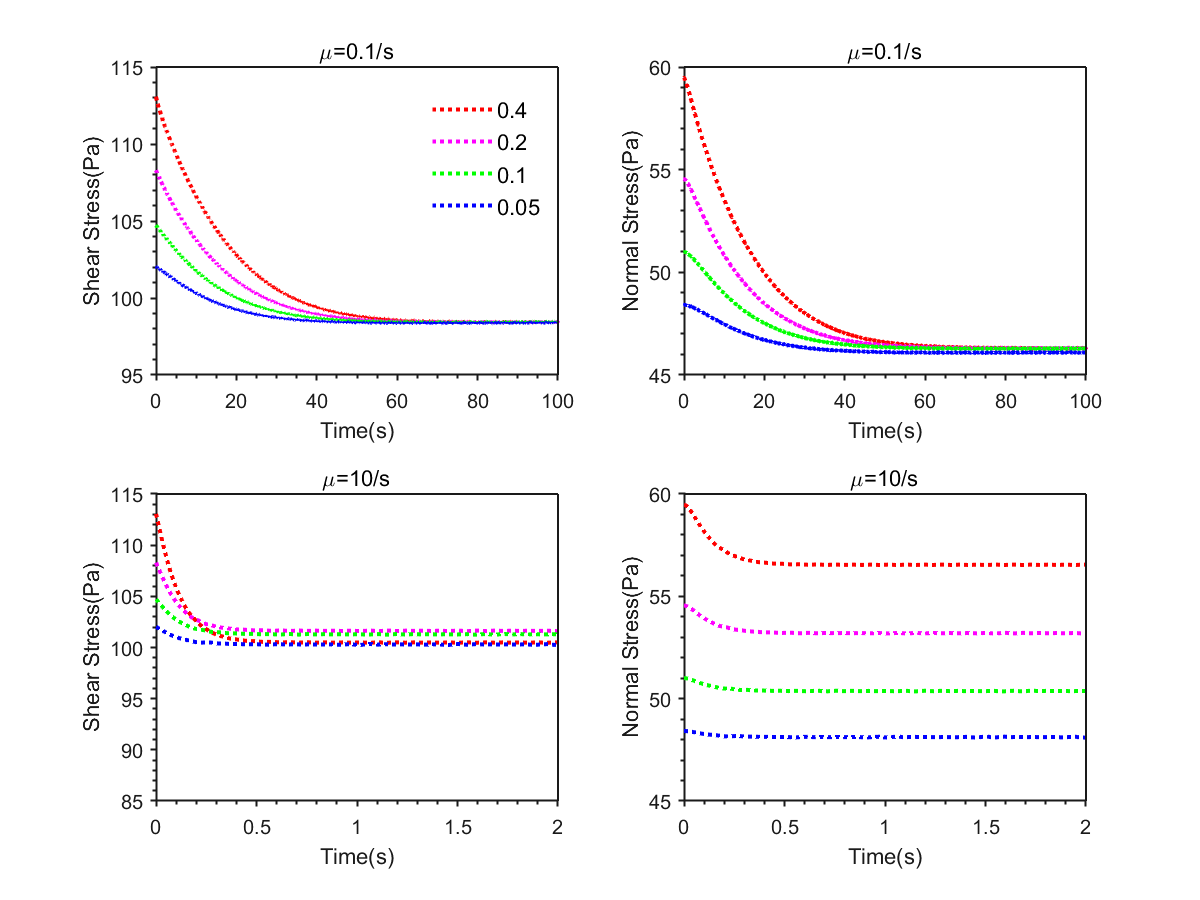}
	\caption[2]{Stress relaxation of the shear and the first normal stress difference of the new TC model with different $\mu$.}
	\label{Stress-relax-2}
\end{figure}

Since the experiment is given in such an idealistic setting. In reality, the shear rate drops to zero in a narrow transition layer. Let's conduct the analysis for this realistic case.
We set the shear rate $\dot{\gamma}(t)$ as a continuous function of time:
\ben
\dot{\gamma}(t)=\left\{
\bea{ll}
\dot{\gamma} & t<0,\\
\dot{\gamma}e^{-\mu t} & t\geq 0.
\eea\right.
\een
Parameter  $\mu$ can be used to control how fast the shear rate decreases. In our model, the results of stress relaxation experiment vary depending on the value of  $\mu$. When $\mu$ is small, indicating the shear rate decreases slowly, the shear stress would converge to the yield stress $\sigma_{y}$ asymptotically, which is different from those from a sudden change in the initial shear rate. When $\mu$ is large, indicating  the shear rate decreases rapidly, the result is similar to the case where the shear rate is discontinuous, namely, the convergent values depend on the initial values and the shear rate drop  rather than the yield stress values. Figure \ref{Stress-relax-2} demonstrates the behavior of the stress relaxation under continuous change in the TC model.

In fact, a simple asymptotic analysis supports the observation. It follows from equation \eqref{Incept-1}-b that the shear stress in large time is approximately given by
\ben
(1-g) \sigma_s\approx \lambda \eta_s \dot{\gamma}-\mu \frac{\lambda \eta_s^2}{G}\dot{\gamma}, \quad \lambda\approx g.
\een
I.e.,
\ben
\sigma_s \approx \sigma_y(1-\frac{\mu \eta_s}{G}).
\een
When, $\mu$ is small, it's close to $\sigma_y$. Otherwise, it's deviate from $\sigma_y$. This explains why the model does not give rises to the yield stress in the sudden withdrawal of the shear.

\subsubsection{Small and large amplitude oscillatory shearing}

\noindent \indent Another rheological experiment is the oscillatory shear(\cite{Rogers2011,Kim2014,Larson1999}).
Shear oscillation experiments are commonly used in rheological studies to observe the stress response to  applied strain $\gamma=\gamma_{0}\sin(\omega t)$ in a complex fluid material. The controlling parameters in the experiment are amplitude $\gamma_{0}$ and oscillation frequency $\omega$.
Note that there is a change in the shear rate of the material when strain is applied as $\gamma=\gamma_{0}\sin(\omega t)$, which implies that the initial value of the shear stress $\sigma_{s}^{0}$ of the material may not be 0. In fact,
\beq
\bea{l}
\dot{\gamma}^{+}-\dot{\gamma}^{-}=\omega\gamma_{0},
\quad
\dot{\sigma}_{s}^{+}-\dot{\sigma}_{s}^{-}=\eta_{s}\omega\gamma_{0}.
\eea
\eeq

\begin{figure}[htbp]
		\centering
		\includegraphics[width=1\linewidth]{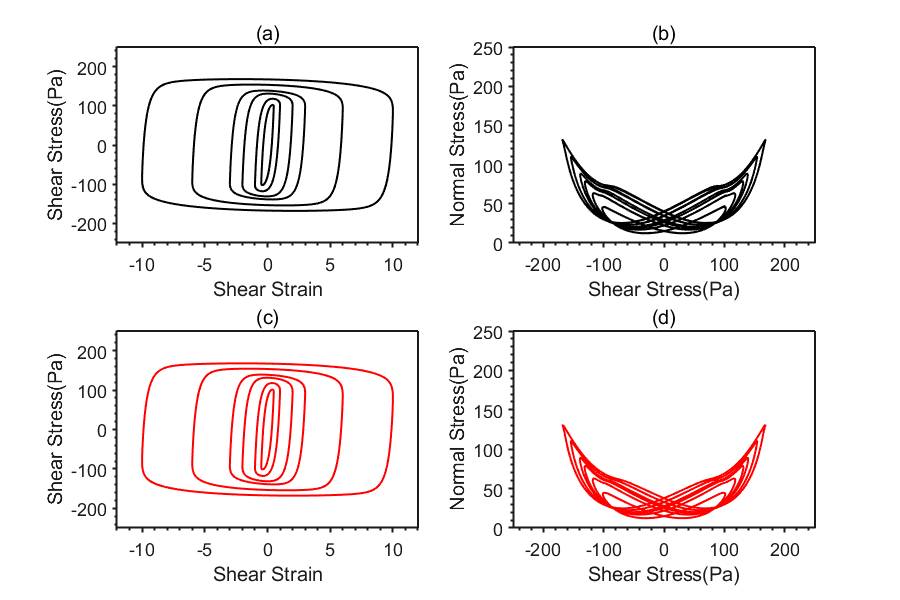}
		\caption[2]{Comparison of the Lissajou curve of the KDR model and the TC model at $\Omega=1rad/s$, where the black curves (a) and (b) are obtained from the KDR model, and the red curves (c) and (d) are from the TC  model. }
\label{Stress-oscil}
\end{figure}

The Lissajous curve of the stress and strain reflects the solid-fluid properties of the material, with the Lissajous curve for solid materials being approximately linear and the Lissajous curve for fluid materials being approximately elliptical. Comparing the curves of shear stress and shear strain of the two models, we observe that the difference between the KDR model and the new TC model is small at small amplitudes, but the difference can be large at large amplitudes, and the peak stress of the KDR model at large amplitudes is smaller than that of the TC model, which is closer to the experimental results \cite{Rogers2021}. However, from the Lissajous curves of the first normal stress difference and shear stress, the new TC model is closer to formula $\sigma_{N}\propto\sigma_{s}^{2}$ reported in \cite{Cagny2018}. Figure \ref{Stress-oscil} depicts the results.

Notice that the two important rheological quantities in oscillatory shear experiments are storage modulus $G^{\prime}$ and loss modulus $G^{\prime\prime}$, which reflect the ability of the material to store and dissipate energy. Under small amplitude shear oscillations, the shear stress $\sigma_{s}$ of the material depends approximately linearly on the shear strain $\gamma$ and shear rate $\dot{\gamma}$, and the corresponding slopes are the storage modulus $G^{\prime}$ and dynamics viscosity $\frac{G^{\prime\prime}}{\omega}$, respectively,
\beq
\sigma_{s}=G^{\prime}\gamma+\frac{G^{\prime\prime}}{\omega}\dot{\gamma}=G^{\prime}\gamma_{0}\sin(\omega t)+G^{\prime\prime}\gamma_{0}\cos(\omega t).
\eeq
In fact, the shear stress also depends on other terms such as $\sin(2\omega t),\cos(2\omega t),\sin(3\omega t),\cos(3\omega t)$, etc. in general shear oscillation experiments without linearization. So it is difficult to obtain the modulus directly from the slope of the Lissajous curve. We can eliminate the high frequency terms by integrating over a period.  The two moduli are then given respectively by
\beq
\bea{l}
G^{\prime}=\frac{2}{T\gamma_{0}^{2}}\int_{t_{0}}^{t_{0}+T}\sigma_{s}\gamma dt,
\quad
G^{\prime\prime}=\frac{1}{\pi\gamma_{0}^{2}}\int_{t_{0}}^{t_{0}+T}\sigma_{s}\dot{\gamma} dt,
\eea
\eeq
where period $T=\frac{2\pi}{\omega}$ and $t_{0}$ is far from  the onset of the shear oscillation.

\begin{figure}[htbp]
	\centering
	\includegraphics[width=1\linewidth]{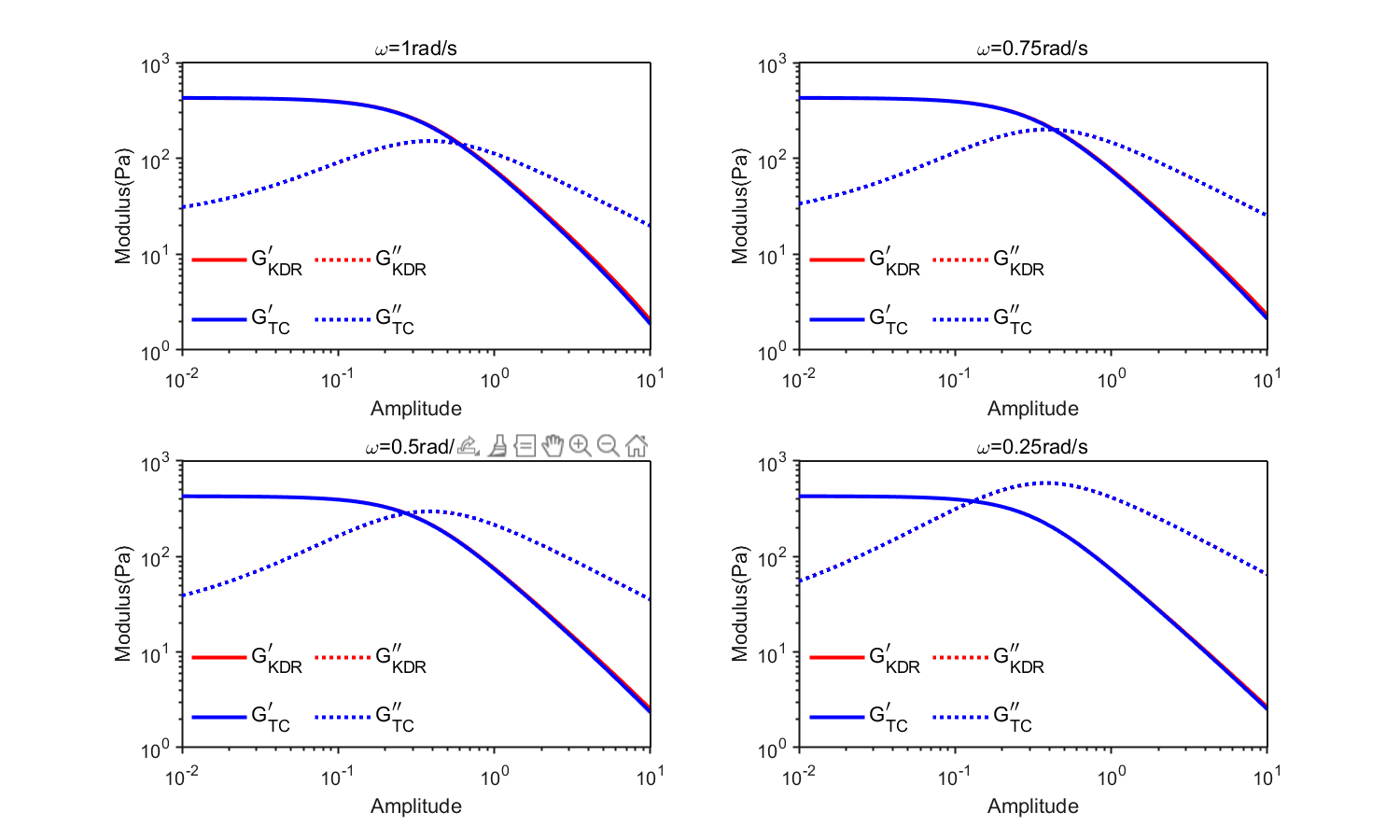}
	\caption[3]{The relationship between the dynamic modulus and amplitude of the KDR model and the new TC model. $G_{R}^{\prime}$ and $G_{R}^{\prime\prime}$ correspond to the storage modulus and loss modulus of the KDR model. $G_{M}^{\prime}$ and $G_{M}^{\prime\prime}$ correspond to the storage modulus and loss modulus of the new TC model.}
	\label{Moduli-1}
\end{figure}
\begin{figure}[htbp]
	\centering
	\includegraphics[width=1\linewidth]{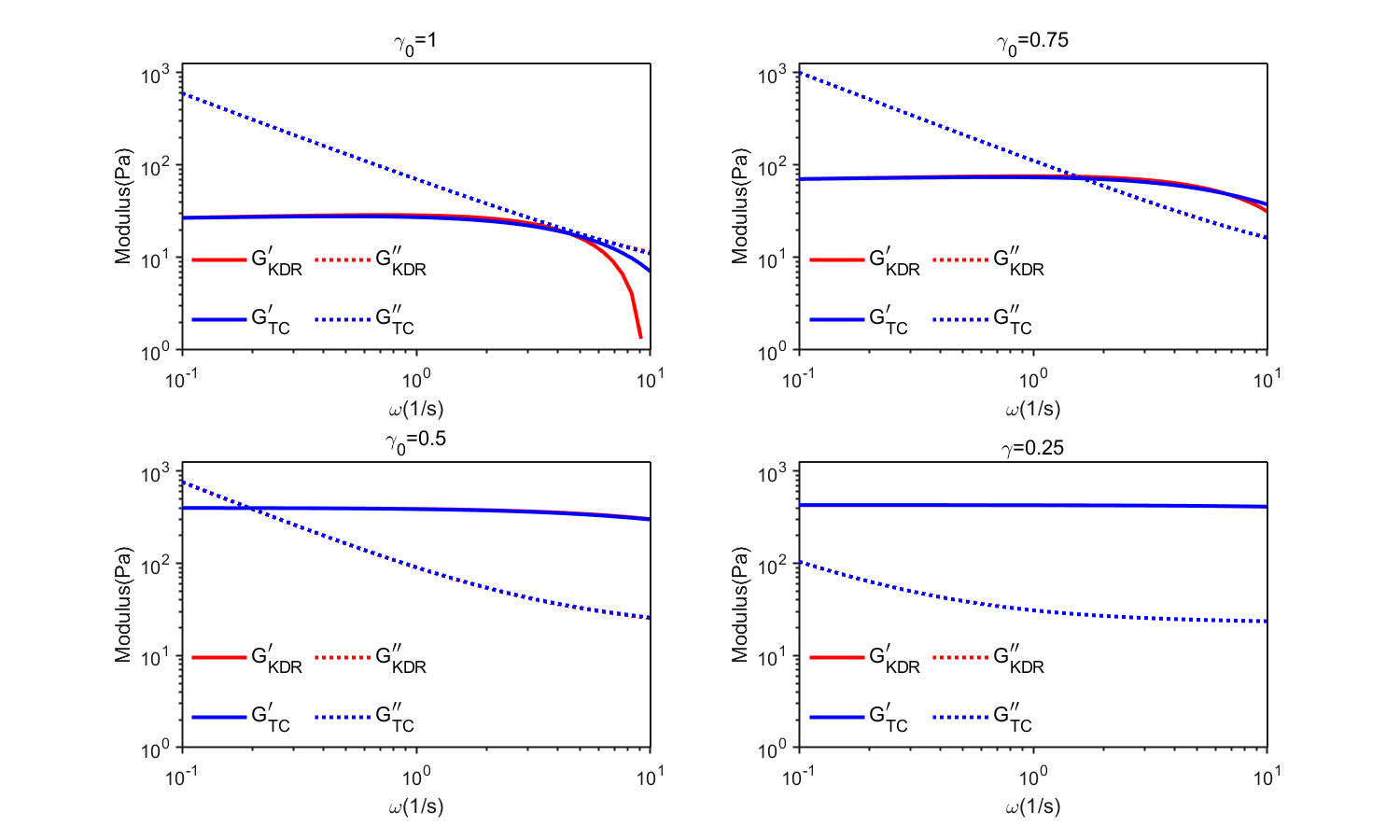}
	\caption[3]{The relationship between the dynamic modulus and frequency of the KDR model and the new TC model. $G_{R}^{\prime}$ and $G_{R}^{\prime\prime}$  correspond to the storage modulus and loss modulus of the KDR model. $G_{M}^{\prime}$ and $G_{M}^{\prime\prime}$ corresponds to the storage modulus and loss modulus of the new TC model.}
	\label{Moduli-2}
\end{figure}

The loss modulus curves for the KDR model and the new TC model exhibit near-identical behavior. However, a slight discrepancy is observed in the storage modulus, despite both models displaying a similar trend at small amplitudes. Notably, this difference becomes more pronounced as the amplitude increases, potentially revealing a divergence between the two models in the nonlinear regime. The storage and loss moduli are depicted in Figures \ref{Moduli-1} and \ref{Moduli-2}, respectively.

\subsubsection{Creep and constrained recoil}

\noindent \indent Creep and constrained recoil experiments are stress-controlled experiments in which the evolution of the deformation and deformation rate are caused by suddenly changing and then maintaining a constant stress value(\cite{Coussot2002,Siebenburger2012,Moller2006,Coussot2006}). It should be noted that the material's deformation is a continuous function of time normally, however, the experiment could lead to discontinuity in the deformation rate because of the sudden change in stress.

Generally, we consider such a shear rate defined by the following constitutive equation for a given stress:
\beq
\sigma_{s}=\sigma_{s}^{-}+(\sigma_{s}^{+}-\sigma_{s}^{-})H(t)=
\bec
\sigma_{s}^{-},\ t<0,\\
\sigma_{s}^{+},\ t\geq0.
\eec
\eeq
We note from the constitutive equation that
\beq
\left \{
\bea{l}
\dot{\gamma}^{-}=
\bec
0,\ \abs{\sigma_{s}^{-}}<\sigma_{y},\\
sgn(\sigma_{s}^{-})(\frac{\abs{\sigma_{s}^{-}}-\sigma_{y}}{k})^{\frac{1}{n}},\ \abs{\sigma_{s}^{-}}\geq\sigma_{y},
\eec,
\\\\
\lambda^{0}=\frac{\sigma_{s}^{-}}{\sigma_{s}^{-}+\eta_{s}\dot{\gamma}^{-}},\\\\
\sigma_{v}^{-}=\frac{\sigma_{s}^{-}\eta_{s}\dot{\gamma}^{-}}{\sigma_{s}^{-}+\eta_{s}\dot{\gamma}^{-}}.
\eea\right.
\eeq
According to  Equation (\ref{mc}), we have
\beq
\bea{l}
\sigma_{v}^{+}=\sigma_{s}^{+}-\frac{(\sigma_{s}^{-})^{2}}{\sigma_{s}^{-}+\eta_{s}\dot{\gamma}^{-}},\quad
\dot{\gamma}^{+}=\frac{\sigma_{s}^{+}}{\eta_{s}}+\frac{\sigma_{s}^{+}}{\eta_{p}^{-}}-\frac{\sigma_{s}^{-}}{\eta_{s}}.
\eea
\eeq

\subsubsection{Creep}

\noindent \indent Creep is an experiment in which the deformation of a material is observed by applying a constant stress value, $\sigma_0$ at $t\geq 0$,
\ben
\sigma_s=
\left\{
\bea{ll}
0 & t<0,\\
\sigma_0 & t\geq 0.
\eea\right.
\een
This experiment can be used to distinguish between the solid and fluid properties of a material, where a solid material resists further deformation after a certain degree of deformation under constant stress while a fluid material will keep flowing.

In creep experiments, the sample changes from static to flowing at a constant shear stress $\sigma_{s}$. The dynamics is described by the following system derived from the constitutive equation:
\beq
\bec
(1-\lambda)\sigma_{N}+\frac{\eta_{s}}{G}\dot{\sigma}_{N}=\frac{2\eta_{s}\dot{\gamma}}{G}(\sigma_{s}-\sigma_{v}),\\
(1-\lambda)\sigma_{s}=\sigma_{v}+\frac{\eta_{s}}{G}\dot{\sigma}_{v},\\
\sigma_{v}=\lambda\eta_{s}\dot{\gamma},\\
\dot{\lambda}=\frac{\alpha G}{\eta_{s}}(g(\dot{\gamma})-\lambda),
\eec
\eeq
with initial  conditions
\ben
\bea{l}
\lambda^{0}=1, \quad
\sigma_{v}^{0}=\sigma_{s}, \quad
\dot{\gamma}^{0}=\frac{\sigma_{s}}{\eta_{s}},\quad
\sigma_{N}^{0}=0.
\eea
\een


Yield stress fluids exhibit a fascinating behavior known as the avalanche phenomenon in creep experiments. When subjected to stress below the yield stress, the material's shear rate diminishes to zero, halting deformation. Conversely, when stress surpasses the yield stress, the shear rate decays to a steady state while the fluid keep flowing. Both the KDR model and the new TC model capture this characteristic, but the KDR model displays a quicker decay in shear rate compared to the new TC model. This discrepancy arises from the dynamic influence of the internal variable $\lambda$, which slows down the shear rate decay. The intriguing creep shear experiments, illustrating selected shear stress values, are depicted in Figure \ref{Creep}.
\begin{figure}[htbp]
	\centering
		\centering
		\includegraphics[width=0.65\linewidth]{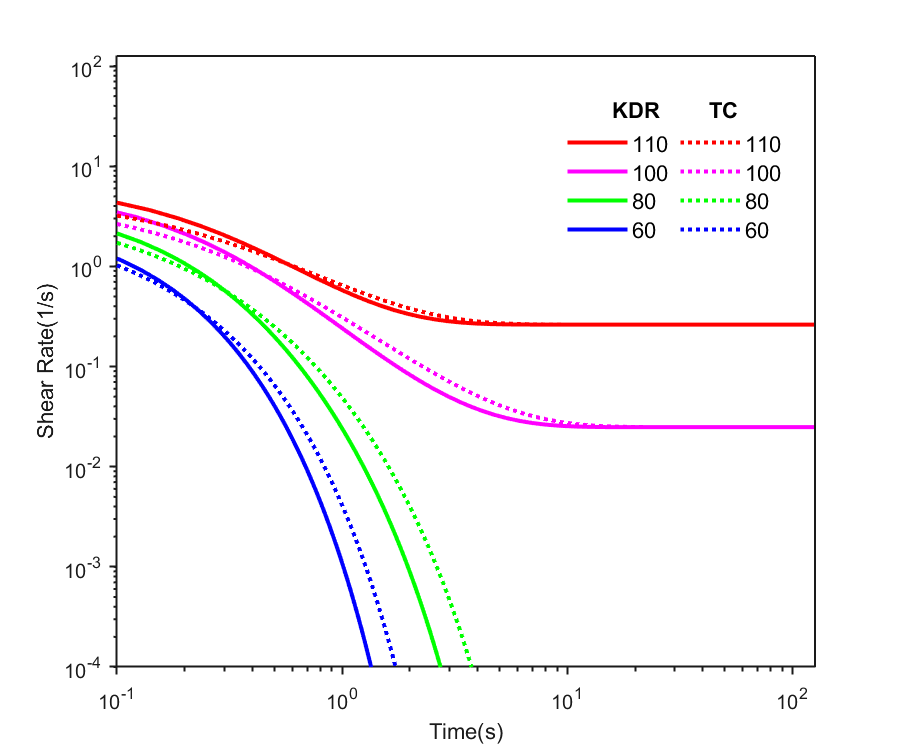}
		\caption[5]{Comparison of creep test results between the KDR model and new TC model at stress values 110, 100, 80 and 60 Pa, respectively. }
\label{Creep}\label{Creep}
\end{figure}

\subsubsection{Constrained recoil after a steady shear flow}

\noindent \indent For a flowing material in steady state shear flow under stress, if one suddenly shuts down the shear stress of the material to 0,
\ben
\sigma_s=
\left\{
\bea{ll}
\sigma_0 & t<0,\\
0 & t\geq 0,
\eea\right.
\een
 one observes that the material will be sheared in the opposite direction relative to the original shear direction and eventually tends to be stationary. This experiment is called recoil. Both the KDR model and the new TC model are capable of  modeling this phenomenon.
In the recoil experiment, the yield stress fluid sample changes abruptly from a steady state shear flow with the  steady shear stress $\sigma_{s}$ to a stress free state. The governing equation system for the relaxation process is given by
\beq
\bec
(1-\lambda)\sigma_{N}+\frac{\eta_{s}}{G}\dot{\sigma}_{N}=-\frac{2\eta_{s}\dot{\gamma}}{G}\sigma_{v},\\
\sigma_{v}+\frac{\eta_{s}}{G}\dot{\sigma}_{v}=0,\\
\sigma_{v}=\lambda\eta_{s}\dot{\gamma},\\
\dot{\lambda}=\frac{\alpha G}{\eta_{s}}(g(\dot{\gamma})-\lambda),
\eec
\eeq
with initial  conditions
\beg
\lambda^{0}=\frac{\sigma_{0}}{\sigma_{0}+\eta_{0}\dot{\gamma}^{-}}, \quad
\sigma_{v}^{0}=-\frac{\sigma_{0}^{2}}{\sigma_{0}+\eta_{s}\dot{\gamma}^{-}}, \quad
\dot{\gamma}^{0}=-\frac{\sigma_{s}}{\eta_{s}}, \quad
\sigma_{N}^{0}=\frac{2}{G}\sigma_{0}^{2}.
\eeg

\begin{figure}[htbp]
	\centering
		\centering
		\includegraphics[width=1\linewidth]{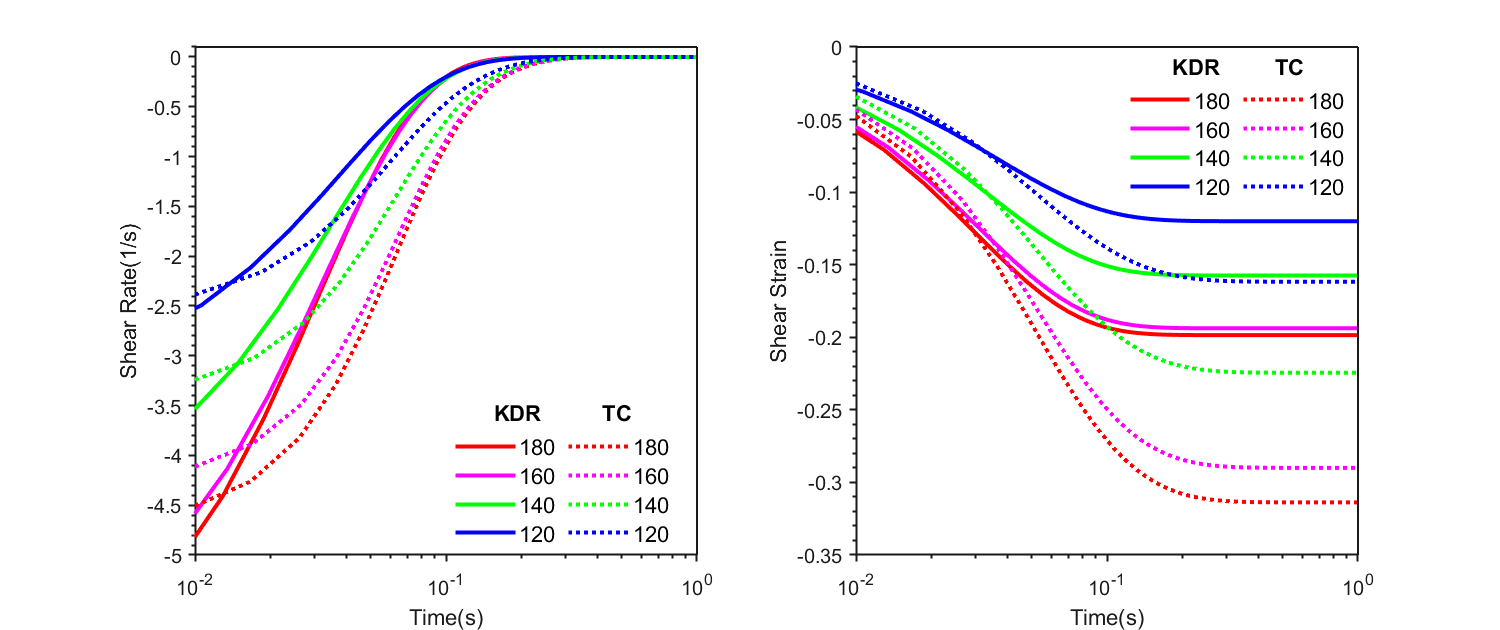}
		\caption[5]{Comparison of recoil test results between the KDR model and the new TC model at stress values 110, 100, 80 and 60 Pa, respectively.}
\label{Recoil}
\end{figure}
We find that the new TC model has a slower initial recoil shear rate but deeper strain recoil compared to those of the KDR model. The comparative results are shown in Figure \ref{Recoil}.

\subsubsection{Stress relaxation after a sudden strain}

\noindent \indent In the stress relaxation experiment of the step steady state flow, the shear strain is assumed continuous while the shear rate is discontinuous but bounded, from which we observe an abrupt change of shear stress dependent on the shear rate difference. Here,  we apply a discontinuous shear strain to measure the stress relaxation after the sudden change in the strain.
Namely, we set
\ben
\gamma=\gamma_{0}H(t), \quad  \dot{\gamma}=\gamma_{0}\delta(t).
\een
Obviously we have $\lambda=1$ for $t>0$.
The governing equations of the dynamics are given by
\beq
\bea{l}
\dot{\sigma}_{N}=2\dot{\gamma}(\sigma_{s}-\eta_{s}\dot{\gamma}),\quad
\dot{\sigma}_{s}=G\dot{\gamma}+\eta_{s}\ddot{\gamma}.
\eea
\eeq
The solutions of the equations are
\beq
\bea{l}
\sigma_{N}=G\gamma^{2}=G\gamma_{0}^{2}H^{2}(t),\quad
\sigma_{s}=G\gamma+\eta_{s}\dot{\gamma}=G\gamma_{0}H(t)+\eta_{s}\gamma_{0}\delta(t).
\eea
\eeq
The shear stress and the first normal stress difference are constants after the sudden strain for the new TC model analogous to the results of the KDR model.

\subsection{Elongational flow}

\noindent \indent  Another important  flow for validating rheological models is a shearfree flow, also known as the elongational flow(\cite{Renardy2009,Daivids2003,James2016}). We consider an elongational flow in the 3D Cartesian coordinate with elongation rate  $\dot{\epsilon}$. The velocity gradient tensor in the coordinate is
given by
\beq
\bL=\frac{\dot{\epsilon}}{2}\begin{pmatrix}
	-1&0&0\\
	0&-1&0\\
	0&0&2
\end{pmatrix}.
\eeq
\beq\dot{\gamma}=\sqrt{3}\dot{\epsilon}\eeq
The stress constitutive equation of (\ref{xxx}) in the componentwise form is given by
\beq
\bec (\frac{1}{\tau}-\frac{\dot{M}}{M})\sigma_{xx}+\dot{\sigma}_{xx}+\dot{\epsilon}\sigma_{xx}=-[(\frac{1}{\tau}-\frac{\dot{M}}{M})\eta+M+\dot{\eta}]\dot{\epsilon}-\eta(\ddot{\epsilon}+\dot{\epsilon}^{2}),\\
    (\frac{1}{\tau}-\frac{\dot{M}}{M})\sigma_{yy}+\dot{\sigma}_{yy}+\dot{\epsilon}\sigma_{yy}=-[(\frac{1}{\tau}-\frac{\dot{M}}{M})\eta+M+\dot{\eta}]\dot{\epsilon}-\eta(\ddot{\epsilon}+\dot{\epsilon}^{2}),\\
	(\frac{1}{\tau}-\frac{\dot{M}}{M})\sigma_{zz}+\dot{\sigma}_{zz}-2\dot{\epsilon}\sigma_{zz}=2[(\frac{1}{\tau}-\frac{\dot{M}}{M})\eta+M+\dot{\eta}]\dot{\epsilon}+2\eta(\ddot{\epsilon}-2\dot{\epsilon}^{2}),\\
	(\frac{1}{\tau}-\frac{\dot{M}}{M})\sigma_{xy}+\dot{\sigma}_{xy}-\dot{\epsilon}\sigma_{xy}=0,\\
	(\frac{1}{\tau}-\frac{\dot{M}}{M})\sigma_{xz}+\dot{\sigma}_{xy}-\frac{3}{2}\dot{\epsilon}\sigma_{xy}=0,\\
	(\frac{1}{\tau}-\frac{\dot{M}}{M})\sigma_{yz}+\dot{\sigma}_{xy}-\frac{3}{2}\dot{\epsilon}\sigma_{xy}=0.
\eec
\eeq
We examine the first normal stress difference $\sigma_{N}=\sigma_{zz}-\sigma_{xx}=\sigma_{zz}-\sigma_{yy}$ and the elongation stress $\sigma_{zz}$. We note that the second normal stress difference $\sigma_{xx}-\sigma_{yy}=0$. In the elongational flow, shear stress components $\sigma_{xy}=\sigma_{xz}=\sigma_{yz}=0$. The constitutive equation for stress reduces to the following system
\beq
\bec
(\frac{1}{\tau}-\frac{\dot{M}}{M}+\dot{\epsilon})\sigma_{N}+\dot{\sigma}_{N}-3\dot{\epsilon}\sigma_{zz}=3[(\frac{1}{\tau}-\frac{\dot{M}}{M})\eta+M+\dot{\eta}]\dot{\epsilon}+3\eta\ddot{\epsilon}-3\eta\dot{\epsilon}^{2},\\
(\frac{1}{\tau}-\frac{\dot{M}}{M})\sigma_{zz}+\dot{\sigma}_{zz}-2\dot{\epsilon}\sigma_{zz}=2[(\frac{1}{\tau}-\frac{\dot{M}}{M})\eta+M+\dot{\eta}]\dot{\epsilon}+2\eta\ddot{\epsilon}-4\eta\dot{\epsilon}^{2}.
\eec
\eeq
It implies
\beq
\bec
[(1-\lambda)+\frac{\eta_{s}\dot{\epsilon}}{G}]\sigma_{N}+\frac{\eta_{s}}{G}(\dot{\sigma}_{N}-3\dot{\epsilon}\sigma_{zz})=\sqrt{3}\sigma_{v}+\frac{\sqrt{3}\eta_{s}}{G}(\dot{\sigma}_{v}-\dot{\epsilon}\sigma_{v}),\\
(1-\lambda)\sigma_{zz}+\frac{\eta_{s}}{G}(\dot{\sigma}_{zz}-2\dot{\epsilon}\sigma_{zz})=\frac{2}{\sqrt{3}}\sigma_{v}+\frac{2}{\sqrt{3}}\frac{\eta_{s}}{G}(\dot{\sigma}_{v}-2\dot{\epsilon}\sigma_{v}),\\
\sigma_{v}=\lambda\eta_{s}\dot{\gamma},\\
\dot{\lambda}=\frac{\alpha G}{\eta_{s}}(g(\dot{\gamma})-\lambda).
\eec
\eeq
In elongational flow, the elongation viscosity also known as the Trouton viscosity is defined by
\ben
\eta_N=\frac{\sigma_N}{\dot{\epsilon}}.
\een

When the system approaches to a steady state asymptotically, we obtain
\ben
\left \{
\bea{l}
\lambda^{0}=g(\dot{\gamma}), \quad
\sigma_{v}=\lambda\eta_{s}\dot{\gamma}, \quad
\sigma_{N}=\eta_{N}\dot{\epsilon}, \quad
\sigma_{zz}=\eta_{zz}\dot{\epsilon}\\
\eta_{N}=\frac{(1-g)(1-a)-2a^{2}}{(1-g)(1-g-a)-2a^{2}}\cdot3g\eta_{s},\quad
\eta_{zz}=\frac{1-2a}{1-g-2a}\cdot2g\eta_{s}, \quad a=\frac{\eta_{s}\dot{\epsilon}}{G}.
\eea\right.
\een
We note that the normal stress component, $\sigma_{zz}$, also has the yield stress, given by
\beq
\frac{1}{\sigma_{zz}}=\frac{1}{2}(\frac{1}{\eta_{p}\dot{\epsilon}}-\frac{1}{G-2\eta_{s}\dot{\epsilon}})=\frac{G-\dot{\epsilon}(\eta_p+2\eta_s)}{2\eta_p \dot{\epsilon}(G-2\eta_s \dot{\epsilon})}.
\eeq
The stress $\sigma_{zz}$ decreases with increasing $\dot{\epsilon}$. The yield stress of $\sigma_{zz}$ is
\beq
\sigma^{\prime}_{y}=\lim\limits_{\dot{\epsilon}\to 0}|\sigma_{zz}|=
\bec
\frac{2G\sigma_{y}}{\sqrt{3}G-\sigma_{y}},\ \dot{\epsilon}\to0^{+}\\
\frac{2G\sigma_{y}}{\sqrt{3}G+\sigma_{y}},\ \dot{\epsilon}\to0^{-}
\eec
\eeq
The steady values of the elongation stress and the normal stress are the same as those of the KDR model. But we need to point it out  that both models are invalid when the elongation rate exceeds critical rate $\dot{\epsilon}_{c}$ which is defined by
\beq
(\eta_{p}+2\eta_{s})\dot{\epsilon}_{c}=G.
\eeq
When the material has a finite gap in the strain rate or the first normal stress difference at the moment t = 0, it satisfies
\beq
\left \{
\bea{l}
\sigma^{+}_{N}-\sigma^{-}_{N}=\frac{3}{2}(\sigma^{+}_{zz}-\sigma^{-}_{zz})=\sqrt{3}(\sigma^{+}_{v}-\sigma^{-}_{v})\\\\
\sigma^{+}_{v}-\sigma^{-}_{v}=g(\dot{\gamma}^{-})\eta_{s}(\dot{\gamma}^{+}-\dot{\gamma}^{-})
\eea\right.
\eeq
In the KDR model, it is given by
\beq
\sigma^{+}_{N}-\sigma^{-}_{N}=\frac{3}{2}(\sigma^{+}_{zz}-\sigma^{-}_{zz})=\int_{\dot{\epsilon}^{-}}^{\dot{\epsilon}^{+}}\frac{3\eta_{s}\eta_{p}}{\eta_{p}+\eta_{s}}d\dot{\epsilon}.
\eeq

\subsubsection{Stress growth}

\noindent \indent In a stress growth experiment, the sample changes abruptly from static to flowing at a constant elongation rate $\dot{\epsilon}$:
\ben
\dot{\epsilon}(t)=
\left \{
\bea{ll}
0 & t<0,\\
\dot{\epsilon} & t\geq 0.
\eea\right.
\een
We calculate the relevant stress response from the following dynamical system:
\beq
\bec
[(1-\lambda)+\frac{\eta_{s}\dot{\epsilon}}{G}]\sigma_{N}+\frac{\eta_{s}}{G}(\dot{\sigma}_{N}-3\dot{\epsilon}\sigma_{zz})=\sqrt{3}\sigma_{v}+\frac{\sqrt{3}\eta_{s}}{G}(\dot{\sigma}_{v}-\dot{\epsilon}\sigma_{v}),\\
(1-\lambda)\sigma_{zz}+\frac{\eta_{s}}{G}(\dot{\sigma}_{zz}-2\dot{\epsilon}\sigma_{zz})=\frac{2}{\sqrt{3}}\sigma_{v}+\frac{2}{\sqrt{3}}\frac{\eta_{s}}{G}(\dot{\sigma}_{v}-2\dot{\epsilon}\sigma_{v}),\\
\sigma_{v}=\lambda\eta_{s}\dot{\gamma},\\
\dot{\lambda}=\frac{\alpha G}{\eta_{s}}(g(\dot{\gamma})-\lambda),
\eec
\eeq
subject to initial conditions
\beg
\lambda^{0}=1, \quad
\sigma_{v}^{+}=\eta_{s}\dot{\gamma}, \quad
\sigma_{N}^{+}=3\eta_{s}\dot{\epsilon}, \quad
\sigma_{zz}^{+}=2\eta_{s}\dot{\epsilon}.
\eeg

\begin{figure}[htbp]
	\centering
		\centering
		\includegraphics[width=\linewidth]{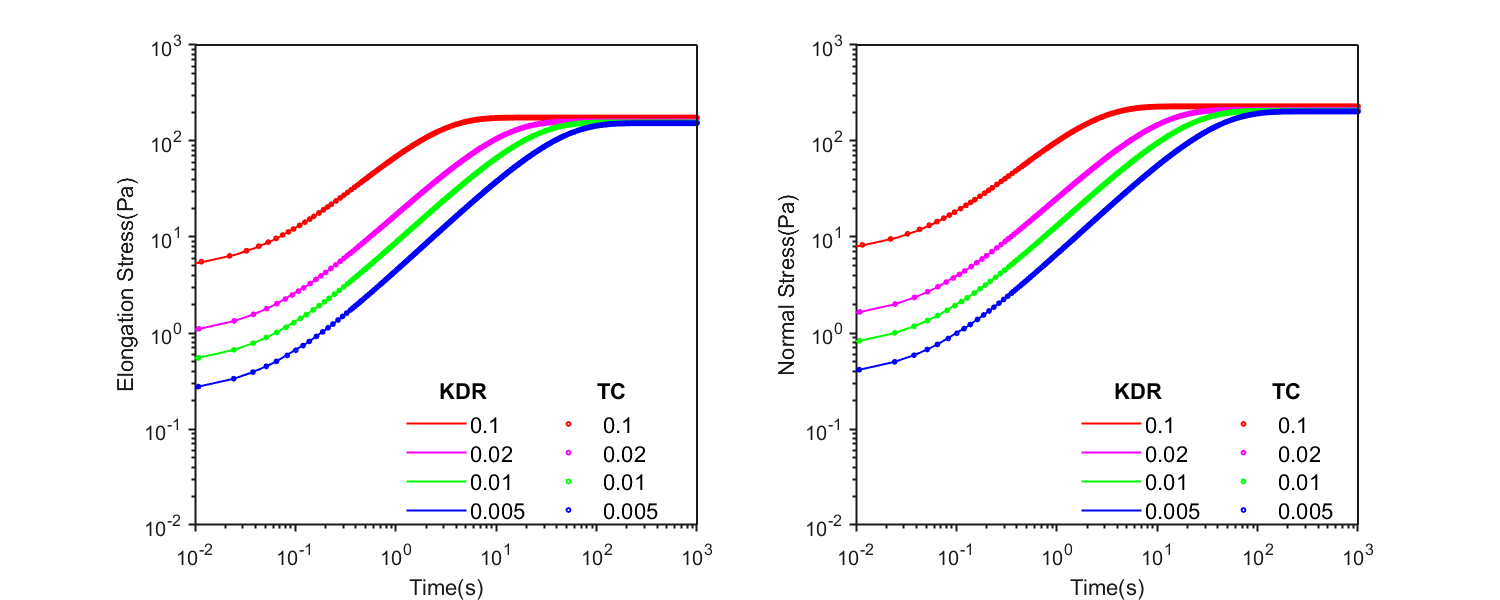}
		\caption[1]{This image shows how the elongation stress and the first normal stress of the KDR model and the new TC model change over time during steady flow at elongation rates of 0.1, 0.02, 0.01, 0.005/s. The solid line represents the KDR model, and the dots represents the modified of the new TC model.
		}\label{Growth-el}
\end{figure}
The elongation stress of the new TC  model agrees well with the KDR model. But the difference in the normal stress difference between the two models is slightly larger in the early stage of stretching and converges thereafter. Unlike in the shear flow, the normal stress associated with the early stage of elongation flow does not scale linearly with the elongation strain. However, the first normal stress difference scales linearly with the elongation rate in the TC model contrary to that of the model of KDR. Figure \ref{Growth-el} depicts the normal stress and the first normal stress differences for a few selected elongation rates.

\subsubsection{Stress relaxation}

\noindent \indent In a stress relaxation experiment, the elongation rate drops abruptly from a nonzero steady elongation rate $\dot{\epsilon}_{0}$ to zero:
\ben
\dot{\epsilon}(t)=\left\{
\bea{ll}
\dot{\epsilon} & t<0,\\
0 & t\geq 0.
\eea\right.
\een

The governing dynamical systems are given by
\beq
\bec
(1-\lambda)\sigma_{N}+\frac{\eta_{s}}{G}\dot{\sigma}_{N}=0,\\
(1-\lambda)\sigma_{zz}+\frac{\eta_{s}}{G}\dot{\sigma}_{zz}=0,\\
\sigma_{v}=0,\\
\dot{\lambda}=\frac{\alpha G}{\eta_{s}}(1-\lambda),
\eec
\eeq
with initial conditions
\beg
\bea{l}
\lambda^{0}=g(\sqrt{3}\dot{\epsilon}_{0})=g, \quad
\sigma_{v}^{0}=0, \\\\
\sigma_{N}^{0}=\frac{3(1-g)g^{2}}{(1-g)(1-g-a)-2a^{2}}\eta_{s}\dot{\epsilon}, \quad
\sigma_{zz}^{0}=\frac{2g^{2}}{1-g-2a}\eta_{s}\dot{\epsilon}.
\eea
\eeg
It follows that
\beq
\frac{d}{dt}(\sigma_{zz}e^{\frac{\lambda}{\alpha}})=\frac{d}{dt}(\sigma_{N}e^{\frac{\lambda}{\alpha}})=0.
\eeq
It indicates that the dynamics of the two rheometric functions are completed determined by the dynamics of the internal variable $\lambda$.
When the system reaches its steady state $\lambda=1$,  we have
\beg
\sigma_{zz}=\frac{2g^{2}}{1-g-2a}\eta_{s}\dot{\epsilon}e^{\frac{1-g}{\alpha}}, \quad
\sigma_{N}=\frac{3(1-g)g^{2}}{(1-g)(1-g-a)-2a^{2}}\eta_{s}\dot{\epsilon}e^{\frac{1-g}{\alpha}}
\eeg

\begin{figure}[htbp]
	\centering
		\centering
		\includegraphics[width=\linewidth]{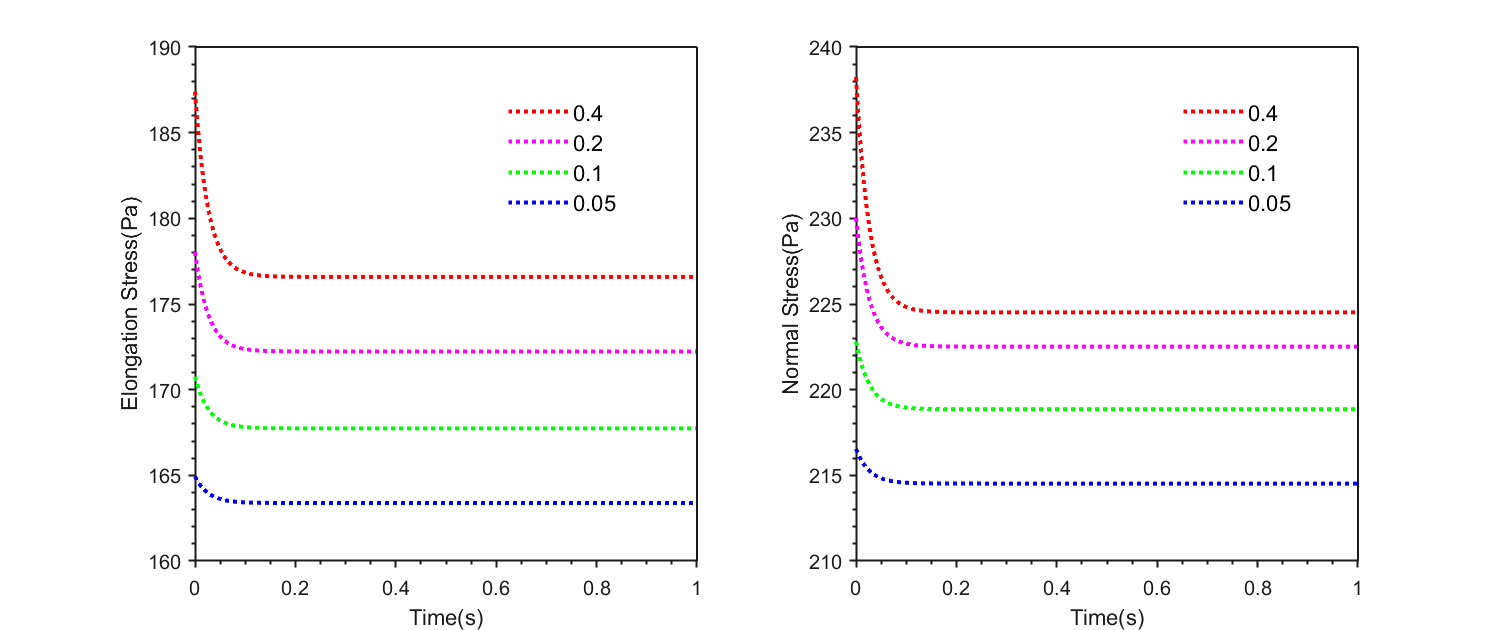}
		\caption[2]{The figure shows how the shear stress and the first normal stress of the new TC model decay after the shear rate from 0.4, 0.2, 0.1, 0.05/s to 0.}\label{Relax-el}
\end{figure}

Similar to the result of the stress relaxation experiment after cessation of a steady shear flow, the elongational stress $\sigma_{zz}$ and the first normal stress difference  $\sigma_{N}$ decay from its initial value to a steady value in the new TC model, whereas the KDR model show that the elongational stress $\sigma_{zz}$ and the first normal stress $\sigma_{N}$ are constants after the elongation rate suddenly changes and given by
\beq
\sigma_{N}^{+}-\sigma_{N}^{-}=\frac{3}{2}(\sigma_{zz}^{+}-\sigma_{zz}^{-})=-\int_{0}^{\dot{\epsilon}_{0}}\frac{3\eta_{p}\eta_{s}}{\eta_{p}+\eta_{s}}d\dot{\epsilon}.
\eeq
Figure \ref{Relax-el} shows the results.

Analogous to the shear experiment, we consider a continuous decay of the elongation by setting
\ben
\dot{\epsilon}=\left\{
\bea{ll}
\dot{\epsilon}_0 & t<0,\\
\dot{\epsilon}_0e^{-\mu t} & t\geq 0.
\eea\right.
\een
The results are shown in the Figure \ref{Relax-el}9. The results of quickly decay are similar to the discontinuous elongation. When the elongation decays slowly, the elongation stress and the first normal stress would converge to their yield values.

\begin{figure}[htbp]
	\centering
	\centering
	\includegraphics[width=\linewidth]{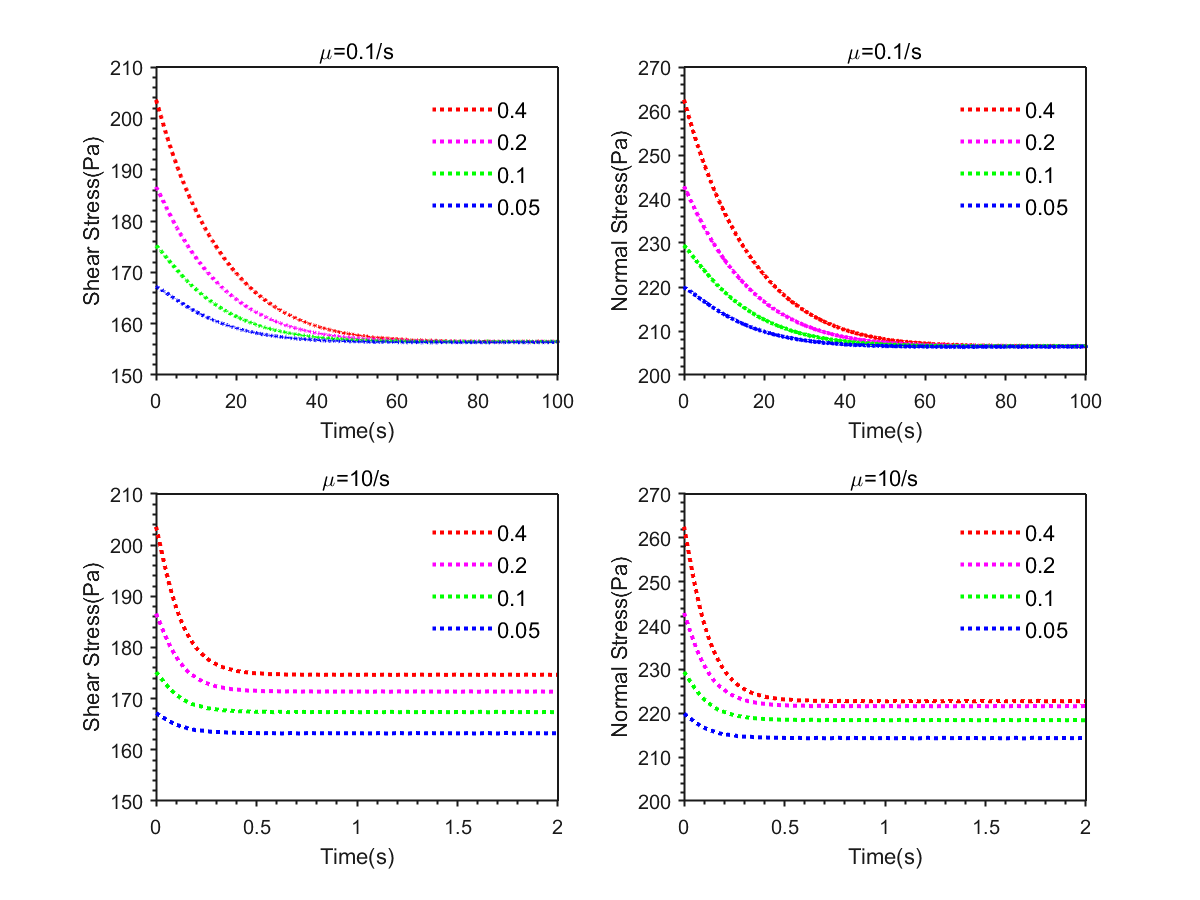}
	\caption[2]{The figure shows how the shear stress and the first normal stress of the new TC model continuous decay from the shear rate from 0.4, 0.2, 0.1, 0.05/s to 0 with different $\mu$ }\label{Relax-el}
\end{figure}

\subsubsection{Creep}

\noindent \indent We apply a constant elongation stress $\sigma_{0}$ at $t\geq 0$
\ben
\sigma_{zz}(t)=
\left\{
\bea{ll}
0 & t<0,\\
\sigma_0 & t\geq 0.
\eea\right.
\een
The dynamics is given by
\beq
\bec
[(1-\lambda)+\frac{\eta_{s}\dot{\epsilon}}{G}]\sigma_{N}+\frac{\eta_{s}}{G}(\dot{\sigma}_{N}-3\dot{\epsilon}\sigma_{zz})=\sqrt{3}\sigma_{v}+\frac{\sqrt{3}\eta_{s}}{G}(\dot{\sigma}_{v}-\dot{\epsilon}\sigma_{v}),\\
(1-\lambda-\frac{2\eta_{s}\epsilon}{G})\sigma_{zz}=\frac{2}{\sqrt{3}}\sigma_{v}+\frac{2}{\sqrt{3}}\frac{\eta_{s}}{G}(\dot{\sigma}_{v}-2\dot{\epsilon}\sigma_{v}),\\
\sigma_{v}=\lambda\eta_{s}\dot{\gamma},\\
\dot{\lambda}=\frac{\alpha G}{\eta_{s}}(g(\dot{\gamma})-\lambda),
\eec
\eeq
with initial  conditions
\beg
\lambda^{0}=1, \quad
\sigma_{v}^{+}=\frac{\sqrt{3}}{2}\sigma_{0}, \quad
\dot{\epsilon}^{+}=\frac{\sigma_{0}}{2\eta_{s}} \quad
\sigma_{N}^{+}=\frac{3}{2}\sigma_{0}.
\eeg
\begin{figure}[htbp]
	\centering
		\centering
		\includegraphics[width=0.65\linewidth]{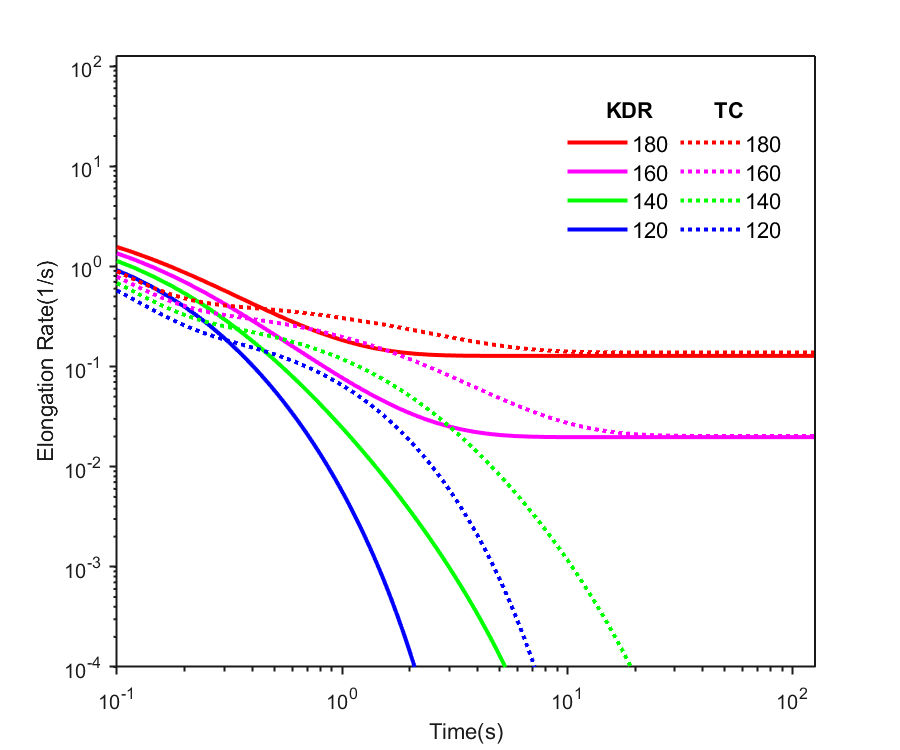}
		\caption[5]{Comparison of creep test results between KDR model and modified Oldroyd-B model when the elongation stress are 180, 160, 140 and 120 Pa}\label{Creep-el}
\end{figure}
Creep results of elongation experiments are similar to shear experiments' that avalanche behavior of strain rate would occur. And the KDR model decays faster than the new TC model under the same elongation stress. The results are shown in Figure \ref{Creep-el}

\subsubsection{Recoil}

\noindent \indent Similar to the shearing recoil experiment, for a flowing material in steady state elongation flow, if one suddenly shuts down the shear stress of the material to 0,
\ben
\sigma_{zz}(t)=
\left\{
\bea{ll}
\sigma_0 & t<0\\
0 & t\geq 0,
\eea\right.
\een
one observes that the material will be elongated in the opposite direction relative to the original shear direction and eventually tends to be stationary. Both of the KDR model and the new TC model are capable of  describing this phenomenon.
The governing systems are given by
\beq
\bec
[(1-\lambda)+\frac{\eta_{s}\dot{\epsilon}}{G}]\sigma_{N}+\frac{\eta_{s}}{G}(\dot{\sigma}_{N}-3\dot{\epsilon}\sigma_{zz})=\sqrt{3}\sigma_{v}+\frac{\sqrt{3}\eta_{s}}{G}(\dot{\sigma}_{v}-\dot{\epsilon}\sigma_{v}),\\
\frac{2}{\sqrt{3}}\sigma_{v}+\frac{2}{\sqrt{3}}\frac{\eta_{s}}{G}(\dot{\sigma}_{v}-2\dot{\epsilon}\sigma_{v})=0,\\
\sigma_{v}=\lambda\eta_{s}\dot{\gamma},\\
\dot{\lambda}=\frac{\alpha G}{\eta_{s}}(g(\dot{\gamma})-\lambda),
\eec
\eeq
with initial  conditions
\beg
\lambda^{0}=g(\dot{\gamma}^{-})=1-g^{-}, \quad
\sigma_{v}^{+}=\sigma_{v}^{-}-\frac{\sqrt{3}}{2}\sigma_{0}, \\
\dot{\epsilon}^{+}=\dot{\epsilon}^{-}-\frac{\sigma_{0}}{2g^{-}\eta_{s}} \quad
\sigma_{N}^{+}=\sigma_{N}^{-}-\frac{3}{2}\sigma_{0}.
\eeg
Here $\dot{\gamma}^{-},\dot{\epsilon}^{-},g^{-}$ satisfy
\beq
\frac{2(1-2a)g}{1-g-2a}\eta_{s}\dot{\epsilon}^{-}=\sigma_{0}
\eeq
\begin{figure}[htbp]
	\centering
		\centering
		\includegraphics[width=\linewidth]{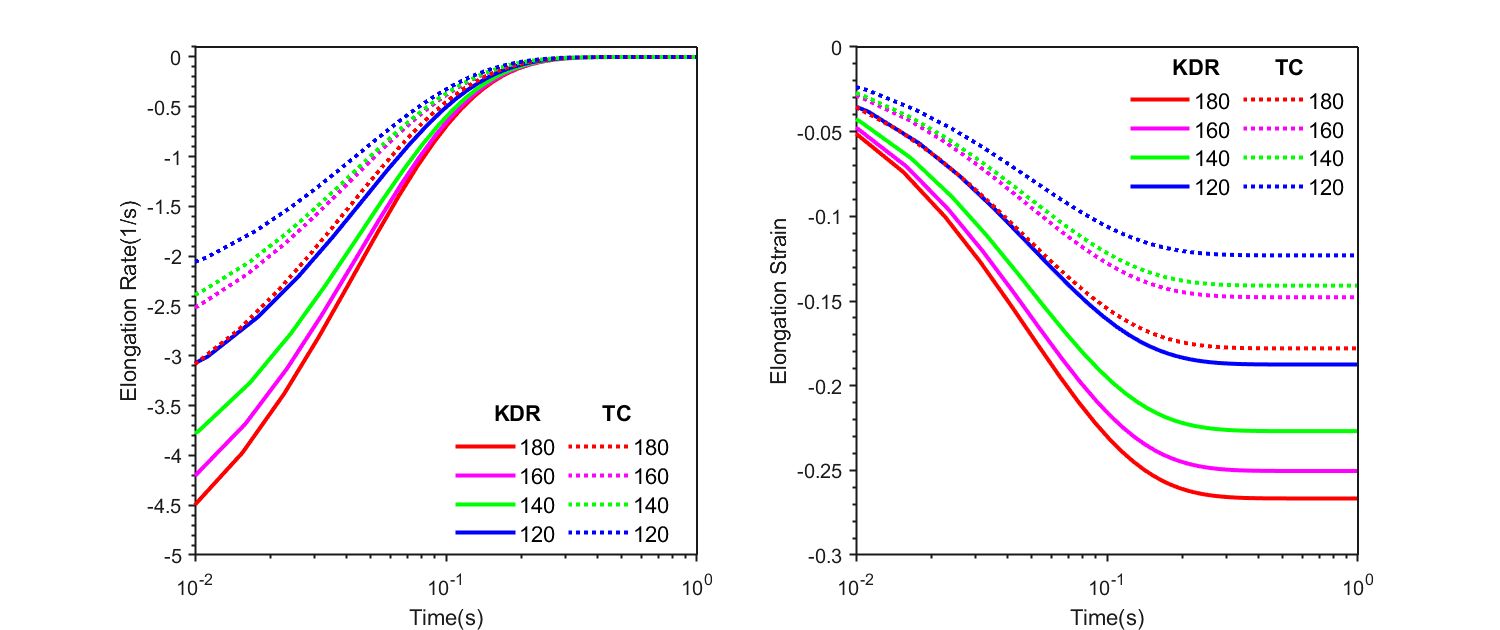}
		\caption[5]{Comparison of recoil test results between the KDR model and the new TC model when the elongation stress are 180, 160, 140 and 120 Pa before the sudden change.}
\end{figure}
In the elongation recoil experiments, The KDR  model consistently stretches more than the TC model and stretches consistently deeper than the TC model

\section{Conclusion}

\noindent \indent

In this study, we have presented a thermodynamically consistent constitutive equation for yield stress fluids by introducing an internal dynamic variable $\lambda$. It is consistent with the Kamani-Donley-Rogers (KDR) model when the parameter $\lambda$ attains a steady-state value in the asymptotic limit. The dynamics of $\lambda$ capture the transient evolution of the mesoscopic structure within the yield stress material.
We compared the numerical results of the new thermodynamically consistent (TC) model and the KDR model under simple shear and elongation flows. The main findings are summarized as follows:
\begin{enumerate}
\item Both models exhibit good agreement in stress growth experiments, although the initial stress values differ.
\item The TC model can capture the decay of shear stress and the first normal stress difference after cessation of shear, while the KDR model holds them constant.
\item The KDR model and the TC model perform comparably at small amplitudes and low frequencies in oscillatory shear experiments.
\item Both models depict transient creep and avalanche behavior, with identical steady-state values. However, the TC model evolves more gradually than the KDR model.
\item Both models can capture recoil behavior. The TC model exhibits more strain recoil compared to the KDR model in shear flows but less in elongation flows.
\end{enumerate}
The advantages of the new TC model over the KDR model lie in its compliance with the second law of thermodynamics and its ability to describe a broader range of rheological properties of yield stress fluids.

\section{Appendix}

\noindent \indent Here, we derive the jump condition between the stress  and the strain rate. We consider a general equation of bounded variables $x, y$ of time $t$ and their time derivatives
\beq
\dot{x}=A(y)\dot{y}+B(x,y)x+C(x,y)y\label{xy}
\eeq
$B(x,y), C(x,y)$ are the bounded scalar functions of $x$ and $y$. We assume that $x, y$ are discontinuous at $t=0$.
\beq
x=x^{-}+(x^{+}-x^{-})H(t)=
\bec
x^{-}(t),\ t<0,\\
x^{+}(t),\ t\geq0,
\eec
\eeq
\beq
y=y^{-}+(y^{+}-y^{-})H(t)=
\bec
y^{-}(t),\ t<0,\\
y^{+}(t),\ t\geq0.
\eec
\eeq
Then the jump condition between $x$ and $y$ at  $t=0$ is given by
\beq
x^{+}(0)-x^{-}(0)=\int_{y^{-}(0)}^{y^{+}(0)}A(y)dy,\label{cc}
\eeq
which is obtained by integrating both sides of  \eqref{xy} over interval $t\in(-\epsilon,\epsilon)$ and then taking limit $\epsilon\to0$.

\subsection{ TC model}
\subsubsection{Shear flow}
The governing equations for the relevant stress components are
\beq
\bec
[\frac{g}{\alpha}+(1-\lambda)(1-\frac{1}{\alpha})]\sigma_{N}+\frac{\eta_{s}}{G}\dot{\sigma}_{N}=\frac{2\sigma_{v}(\sigma_{s}-\sigma_{v})}{\lambda G},\\
[\frac{g}{\alpha}+(1-\lambda)(1-\frac{1}{\alpha})]\sigma_{s}+\frac{\eta_{s}}{G}\dot{\sigma}_{s}=[\lambda+\frac{g}{\alpha}+(1-\lambda)(1-\frac{1}{\alpha})]\sigma_{v}+\frac{\eta_{s}}{G}\dot{\sigma}_{v},\\
\sigma_{v}=\lambda\eta_{s}\dot{\gamma},\\
\dot{\lambda}=\frac{G}{2\eta_{s}}\lambda[1-g(\dot{\gamma})-\lambda],
\eec
\eeq
where $\lambda$, $\dot{\gamma}$, $g(\dot{\gamma})$, $\sigma_{s}$ and $\sigma_{N}$ are bounded.
So we get $\lambda$ and $\sigma_{N}$ are continuous. Using \eqref{cc}, we have
\beq
\sigma_{N}^{+}-\sigma_{N}^{-}=0,
\quad
\sigma_{s}^{+}-\sigma_{s}^{-}=\sigma_{v}^{+}-\sigma_{v}^{-}=\lambda^{-}\eta_{s}(\dot{\gamma}^{+}-\dot{\gamma}^{-}).
\eeq

\subsubsection{Elongational flow}
The governing equations for the relevant stress components in the elongational flow are
\beq
\bec
[\frac{g}{\alpha}+(1-\lambda)(1-\frac{1}{\alpha})+\frac{\eta_{s}\dot{\epsilon}}{G}]\sigma_{N}+\frac{\eta_{s}}{G}(\dot{\sigma}_{N}-3\dot{\epsilon}\sigma_{zz})=\\
\hskip 1 in \sqrt{3}[\lambda+\frac{g}{\alpha}+(1-\lambda)(1-\frac{1}{\alpha})]\sigma_{v}+\frac{\sqrt{3}\eta_{s}}{G}\dot{\sigma}_{v}-\frac{\sigma^{2}_{v}}{\lambda G},\\
[\frac{g}{\alpha}+(1-\lambda)(1-\frac{1}{\alpha})]\sigma_{zz}+\frac{\eta_{s}}{G}(\dot{\sigma}_{zz}-2\dot{\epsilon}\sigma_{zz})=\\
\hskip 1 in \frac{2}{\sqrt{3}}[\lambda+\frac{g}{\alpha}+(1-\lambda)(1-\frac{1}{\alpha})]\sigma_{v}+\frac{2}{\sqrt{3}}\frac{\eta_{s}}{G}\dot{\sigma}_{v}-\frac{4\sigma^{2}_{v}}{3\lambda G},\\
\sigma_{v}=\lambda\eta_{s}\dot{\gamma},\quad
\dot{\lambda}=\frac{G}{2\alpha\eta_{s}}\lambda[1-g(\dot{\gamma})-\lambda],
\eec
\eeq
where $\dot{\epsilon}=\frac{1}{\sqrt{3}}\dot{\gamma}$, $\lambda$, $g(\dot{\gamma})$, $\sigma_{zz}$ and $\sigma_{N}$ are bounded.
Using \eqref{cc}, we have
\beq
\sigma_{v}^{+}-\sigma_{v}^{-}=\frac{1}{\sqrt{3}}(\sigma_{N}^{+}-\sigma_{N}^{-})=\frac{\sqrt{3}}{2}(\sigma_{zz}^{+}-\sigma_{zz}^{-})=
\lambda^{-}\eta_{s}(\dot{\gamma}^{+}-\dot{\gamma}^{-})=\sqrt{3}\lambda^{-}\eta_{s}(\dot{\epsilon}^{+}-\dot{\epsilon}^{-}).
\eeq

\subsection{ KDR model}
\subsubsection{Shear flow}
The governing equations of the relevant stress components are
\beq
\bec
\sigma_{N}+\frac{\eta_{p}+\eta_{s}}{G}\dot{\sigma}_{N}=2(\sigma_{s}-\frac{\eta_{p}\eta_{s}}{\eta_{p}+\eta_{s}})\frac{\eta_{p}+\eta_{s}}{G}\dot{\gamma},\\
\sigma_{s}+\frac{\eta_{p}+\eta_{s}}{G}\dot{\sigma}_{s}=\eta_{p}(\dot{\gamma}+\frac{\eta_{s}}{G}\ddot{\gamma}),
\eec
\eeq
where $\dot{\gamma}$, $\sigma_{s}$ and $\sigma_{N}$ are bounded. Using \eqref{cc}, we have
\beq
\sigma_{N}^{+}-\sigma_{N}^{-}=0,
\quad
\sigma_{s}^{+}-\sigma_{s}^{-}=\int_{\dot{\gamma}^{-}}^{\dot{\gamma}^{+}}\frac{\eta_{p}\eta_{s}}{\eta_{p}+\eta_{s}}d\dot{\gamma}.
\eeq
\subsubsection{Elongational flow}
The governing equations of the relevant stress components are
\beq
\bec
(1+\dot{\epsilon}\frac{\eta_{p}+\eta_{s}}{G})\sigma_{N}+\frac{\eta_{p}+\eta_{s}}{G}(\dot{\sigma}_{N}-3\dot{\epsilon}\sigma_{zz})=
3\eta_{p}(\dot{\epsilon}+\frac{\eta_{s}}{G}\ddot{\epsilon}-\frac{\eta_{s}}{G}\dot{\epsilon}^{2}),\\
(1-2\dot{\epsilon}\frac{\eta_{p}+\eta_{s}}{G})\sigma_{zz}+\frac{\eta_{p}+\eta_{s}}{G}\dot{\sigma}_{zz}=
2\eta_{p}(\dot{\epsilon}+\frac{\eta_{s}}{G}\ddot{\epsilon}-\frac{\eta_{s}}{G}\dot{\epsilon}^{2}),
\eec
\eeq
where $\dot{\gamma}$, $\sigma_{s}$ and $\sigma_{N}$ are bounded. Using \eqref{cc}, we have
\beq
\sigma_{N}^{+}-\sigma_{N}^{-}=\frac{3}{2}(\sigma_{zz}^{+}-\sigma_{zz}^{-})=\int_{\dot{\epsilon}^{-}}^{\dot{\epsilon}^{+}}\frac{3\eta_{p}\eta_{s}}{\eta_{p}+\eta_{s}}d\dot{\epsilon}.
\eeq

\bibliographystyle{plain}
\bibliography{references}

\end{document}